\documentclass[fleqn, usenatbib]{mnras} 

\usepackage{newtxtext,newtxmath}

\usepackage[T1]{fontenc}
\usepackage{ae,aecompl}
\usepackage{gensymb}

\usepackage[normalem]{ulem}

\usepackage{graphicx}	
\usepackage{amsmath}	
\usepackage{amssymb}	
\title[Cygnus X-1 mass measurement]{An X-ray reverberation mass measurement of Cygnus X-1}

\author[Mastroserio et al.
]{
	Guglielmo Mastroserio,$^{1}$\thanks{E-mail: g.mastroserio@uva.nl}
	Adam Ingram$^{2} $ and
	Michiel van der Klis$ ^{1} $
	\\
	$ ^{1} $Astronomical Institute, Anton Pannekoek, Univerity of Amsterdam, Science Park 904, NL-1098 XH Amsterdam, Netherlands\\
	$ ^{2} $Department of Physics, Astrophysics, University of Oxford, Denys Wilkinson Building, Keble Road, Oxford OX1 3RH, UK
}

\date{Accepted XXX. Received YYY; in original form ZZZ}

\pubyear{2019}

\begin{document}
\label{firstpage}
\pagerange{\pageref{firstpage}--\pageref{lastpage}}
\maketitle

\begin{abstract}
We present the first X-ray reverberation mass measurement of a stellar-mass black hole. Accreting stellar-mass and supermassive black holes display characteristic spectral features resulting from reprocessing of hard X-rays by the accretion disc, such as an Fe K$\alpha$ line and a Compton hump. This emission probes of the innermost region of the accretion disc through general relativistic distortions to the line profile. However, these spectral distortions are insensitive to black hole mass, since they depend on disc geometry in units of gravitational radii. Measuring the reverberation lag resulting from the difference in path length between direct and reflected emission calibrates the absolute length of the gravitational radius. We use a relativistic model able to reproduce the behaviour of the lags as a function of energy for a wide range of variability timescales, addressing both the reverberation lags on short timescales and the intrinsic hard lags on longer timescales.  We jointly fit the time-averaged spectrum and the real and imaginary parts of the cross-spectrum as a function of energy for a range of Fourier frequencies to \textit{Rossi X-ray Timing Exporer} data from the X-ray binary Cygnus X-1. We also show that introducing a self-consistently calculated radial ionisation profile in the disc improves the fit, but requires us to impose an upper limit on ionisation profile peak to allow a plausible value of the accretion disc density. This limit leads to a mass value more consistent with the existing dynamical measurement.
\end{abstract}




\section{Introduction}
Black hole X-ray binary systems radiate a large X-ray flux due to accretion of matter from the companion star onto the black hole.
When active, these sources
display transitions between a hard state, when the hard ($\gtrsim 3$ keV) X-ray flux in the energy 
spectrum is higher than the soft X-ray flux, and a soft state when the situation is the opposite. 
The soft emission is dominated by radiation from the accretion disc, which is modelled with a multi-temperature black body \citep{Shakura1973}. 
The hard component of the spectrum is due to inverse-Compton emission, most likely coming from a cloud of hot electrons 
close to the black hole, often referred to as the `corona' \citep{Thorne1975}. 
Radiation originating from the corona illuminates the accretion disc, where it is  
re-processed and then re-emitted. The resulting `reflection' spectrum includes characteristic features such as a prominent iron K$\alpha$ fluorescence line at $\sim 6.4$ keV
and a Compton hump peaking at $\sim 20-30$ keV \citep[e.g.][]{George1991,Ross2005,Garcia2010, Fabian2010}. 
Modelling the observed reflection spectrum provides a powerful tool for measuring the system parameters, because the emission is distorted by the strong gravity and rapid orbital motion of material close to the compact object. 
Whereas the restframe reflection spectrum depends on properties of the accretion disc such as ionisation state, the gravitational distortions depend strongly on geometrical parameters such as the disc inner radius and inclination angle. Time-averaged spectral analysis alone, however, is not at all sensitive to black hole mass, because the gravitational distortions depend on distances in units of gravitational radii ($R_g=GM/c^2$). 

Besides the long term variability characterizing the state changes, accreting stellar mass black holes show rapid variability in the range 
of milliseconds to tens of seconds. Since the rays that reaches us via reflection follow a longer path length than those we observe directly, fluctuations in the inverse-Compton emission should be followed after a light-crossing delay by similar fluctuations in the reflected emission. The reflection signal is not only lagged with respect to the direct signal, but also smeared such that the fastest variability is washed out. This is because reflection from different parts of the disc is associated with different path lengths, such that a very short flash of X-rays from the corona would result in an extended flare of reflected emission. Whereas the spectral distortions depend on distances in mass scaled units, the reverberation lag between direct and reflected emission depends on distances in absolute units. Therefore measurement of a reverberation lag can calibrate the length of the gravitational radius for a system, providing a way to measure black hole mass \citep{Stella1990} that is orthogonal to other methods \citep[e.g.][and references therein]{Casares2014}.

A combined spectral and timing analysis can be achieved by calculating cross spectra between light curves in different energy bands (e.g. \citealt{Klis1987, Nowak1999}). This can be used to calculate time lag vs energy spectra for different timescale ranges \citep{Mastroserio2018}.
The lag spectrum of long timescale variability (i.e. Fourier frequencies $\nu \lesssim 300 M_\odot/M$ Hz) is observed both in X-ray binaries and active galactic nuclei (AGN) to be featureless, with hard photons lagging soft photons \citep{Miyamoto1988,Nowak1999,Papadakis2001,McHardy2004}. These hard lags, which are far larger than the expected reverberation lags, are likely due to spectral variability of the directly observed coronal emission and they have been attributed to propagating mass accretion rate fluctuations in the accretion flow \citep{Lyubarskii1997, Kotov2001, Arevalo2006,Ingram2013}. In models considering an extended corona, fluctuations propagate from the soft X-ray emitting region further from the black hole to the hard X-ray emitting region closer to the black hole, giving rise to the hard lags. In alternative models considering a compact corona, propagating fluctuations in the disc instead cause variable heating and cooling of the corona, leading to hard lags driven by spectral pivoting of the inverse-Compton spectrum (Uttley \& Malzac in prep). 
The propagating fluctuations mechanism has successfully explained the spectral-timing properties of a number of black hole X-ray binaries \citep{Kotov2001}, although questions remain \citep{Rapisarda2016,Rapisarda2017b,Veledina2018}.

The magnitude of the hard lags is observed to reduce with Fourier frequency, offering the possibility to detect reverberation signatures at high frequencies ($\nu \gtrsim 300 M_\odot/M$ Hz).
Such signatures were first detected for AGN, first in the form of soft reflected X-rays lagging hard directly observed X-rays \citep{Fabian2009} and later in the form of an iron K$\alpha$ feature in the lag-energy spectrum \citep{Zoghbi2011,Kara2016}.
Although alternative models have been proposed for the soft lags \citep[e.g.][]{Miller2010,Mizumoto2018}, the iron K feature seems to provide clear evidence of reverberation. 
Detection of reverberation signatures has been more challenging in the case of Galactic black hole systems, owing to the much shorter associated light-crossing timescale. Soft lags attributed to thermally reprocessed photons lagging directly observed hard photons (thermal reverberation) were the first signature to be observed \citep{Uttley2011,DeMarco2015}. Recently, \citet{Kara2019} reported on the first significant detection of iron K alpha lags, using \textit{NICER} data from MAXI~J1820$+$070.

Still further information is contained in the frequency and energy dependent correlated variability amplitude, which can also be measured using the cross-spectrum. Modelling this in addition to the time-averaged spectrum and the frequency dependent lag energy spectrum can yield a better constraint on black hole mass.
However, in order to probe the frequency dependence of the reverberation lags, 
we must additionally take account of the hard lags. In \citet{Mastroserio2018} 
we developed a formalism to model what we termed the \textit{complex covariance} over a wide range of Fourier frequencies.
We represented the observed hard lags in the coronal emission by a time dependent perturbation
in the illuminating spectrum, and then self-consistently calculated 
the reverberation lags using a transfer function formalism \citep[e.g.][]{Reynolds1999, Wilkins2013}. 
The formalism provides a solid mathematical framework to address 
the non-linear effects in the reverberation introduced by spectral 
changes in the coronal emission.
 
However, the analysis we presented in \citet{Mastroserio2018} did not include the effect of gravitational light bending, even though our best fitting model featured emission from very close to the black hole. 
In Ingram et al (2019) we address this by incorporating a fully general-relativistic ray-tracing calculation for the transfer function, but without considering the non-linear effects. 
Here, for the first time, we include both the non-linear effects and general relativistic effects. We apply our model to \textit{RXTE} data from Cygnus X-1 in order to place constraints on the mass of the central black hole. We explore a number of different assumptions for the radial dependence of the disc ionization parameter, and find that the mass measurement is sensitive to which assumption we make.

In Section~\ref{sec:Methods} we discuss how the non-linear effects of the complex covariance model are affected by the light bending calculation.
In Section~\ref{sec:data} we describe the data and in Section~\ref{sec:en_spectrum} we present the fits to the time averaged spectrum using models with three different ionisation profiles of the accretion disc. 
In Section~\ref{sec:rev_mapping} we perform joint fits to the complex covariance in 10 different frequency ranges and the time averaged spectrum. In Section~\ref{sec:discussion} we discuss our results.

\section{The non-linear model}
\label{sec:Methods}
We model the X-ray corona as a stationary point source located on the disc rotation axis at height $h$ above the black hole (the lamppost model) and the disc as flat and geometrically thin, with its angular momentum axis aligned with the spin axis of the black hole. 
We consider two components of the X-ray spectrum. The first is the directly observed radiation from the point source (`continuum') and the second is radiation that has been re-processed and re-emitted by the disc before reaching the observer (`reflection').
Both of these spectral components are distorted by the strong gravitational field in the vicinity of the black hole.
We calculate the time-dependent energy spectrum observed by a distant observer by starting with the equations presented in Ingram et al (2019) and introducing non-linear effects due to fluctuations in the slope of the continuum spectrum.
In the following sub-sections we study the two components separately before combining them to provide an expression for the total radiation seen by the distant observer.

\subsection{Direct Emission}
We choose an exponentially cut-off power-law function to reproduce the shape of the inverse-Compton emission from the corona. 
Following Ingram et al. (2019) the directly observed specific flux seen by the distant observer is 
\begin{equation}
\begin{split}
F_{o} \left(E_o,t\right)=A(t')\, l\, g_{so} ^{\Gamma-\beta\left(t'\right)}   E_o^{1-\Gamma+\beta\left(t'\right)} e^{-E_o/(g_{so}E_{cut})}  
\end{split}
\label{eq:continuum_non_lin}
\end{equation}
where $A$ is the normalisation, $\Gamma$ and  $E_{cut}$ are respectively the power-law index and the high energy cut-off 
(the latter is related to the temperature of the corona) 
and $l$ is the lensing factor due to light bending (see Ingram et al 2019, equation 10). $g_{so}= E_o/E_s$ is the blueshift experienced by photons travelling from the source to the observer, $E_o$ and $E_s$ are photon energy measured respectively in the observer and source restframe. 
Here, we have introduced fluctuations in the slope of the spectrum through the function $\beta(t)$, 
which depends on the time taken for the photons to travel from the source to the observer, $\tau_{so}$, as $t'= t-\tau_{so}$.  Eq.~\ref{eq:continuum_non_lin} is non-linear in time, though we can linearise the expression by Taylor expanding to the first order around $\beta = 0$ \citep[see for details][]{Mastroserio2018}. The observed specific flux becomes  
\begin{equation}
\begin{split}
F_{o} &\left(E_o,t\right)\approx  l\,g_{so}^{\Gamma}\, E_o^{1-\Gamma} e^{-E_o/\left(g_{so}E_{cut}\right)}   \\ 
&\left[A\left(t'\right)+B\left(t'\right) (\ln E_o -\ln g_{so} )\right]
\end{split}
\label{eq:final_continuum}
\end{equation}
where we have defined $B(t) = A(t) \beta(t)$.  For brevity let us define $D(E) =l\,g_{so}^{\Gamma}\, E^{1-\Gamma} e^{-E/\left(g_{so}E_{cut}\right)} $.

\subsection{Re-processed emission}
Some of the photons emitted from the corona illuminate the accretion disc. They are re-processed and radiated an-isotropically. 
Again following Ingram et al. (2019) we can derive the expression for the reflected specific flux seen by the distant observer emitted from a patch of the disc which subtends a solid angle $d\Omega_d$ in the source restframe and has a surface area $dA_d$ in its own rest frame.
A disc patch with coordinates $(r,\phi)$ subtends a solid angle on the observer plane of $d\Omega= d\alpha_0\, d\beta_0/D^2 $ where $\alpha_0$ and $\beta_0$ are the impact parameters at infinity (\citealt{Luminet1979}). The reflected flux from the disc patch is then
\begin{equation}
\begin{split}
dR_o\left(E_o,t|\mu_e,r,\phi\right) &= A\left(t''\right) g_{do}^{3} \epsilon(r,t'') \times \\
 &\mathcal{R}\left(\frac{E_o}{g_{do}}|\Gamma-\beta(t''),g_{sd}E_{cut} \right)  d\alpha_0 \,d\beta_0
\end{split}
\label{eq:reflection_specific_flux_disc_patch}
\end{equation}
where  
\begin{equation}
 \epsilon(r,t) = \frac{g_{sd}^{\Gamma -\beta(t)}}{4\pi} \frac{d\Omega_d}{dA_d}.
\end{equation}
is the radial emissivity profile and the two factors $g_{sd}(r)$ and $g_{do}(r,\phi)$ represent the blue-shift of the photons travelling from the source to the disc and from the disc to the observer respectively.
$\mathcal{R}$ is the restframe reflection spectrum which depends on the incident radiation, $\mu_e$ is the cosine of the angle between the disc normal and the emergent trajectory of the reflected rays from the plane of the disc and $\beta$ depends on the time taken from the photons to travel from the source to the disc, $\tau$, as $t'' = t - \tau$.  
Note that oscillations in the illuminating power-law slope introduce non-linear effects in the emissivity profile \textit{and} in the restframe reflection spectrum. We can again linearise Eq.~\ref{eq:reflection_specific_flux_disc_patch} by Taylor expanding around $\beta=0$:
\begin{equation}
dR(E,t) \approx dR(E|\beta=0)+ \beta(t) \frac{\partial (dR)}{\partial \beta}(E|\beta=0).
\label{eq:reflection_taylor_expansion}
\end{equation}
Considering $\partial\mathcal{R}/\partial\beta= - \partial\mathcal{R}/\partial\Gamma$,  Eq.~\ref{eq:reflection_specific_flux_disc_patch} becomes 
\begin{equation}
\begin{split}
dR_o\left(E_o,t|r,\phi \right) &=  g_{do}^{3} \epsilon(r,\beta=0) \bigg[ A\left(t-\tau\right) \mathcal{R}\left(\frac{E_o}{g_{do}}|\beta=0 \right)  \\
 - &B\left(t-\tau \right) \mathcal{R}\left(\frac{E_o}{g_{do}}|\beta=0 \right) \ln g_{sd} \\
  - &B\left(t-\tau \right)\frac{\partial\mathcal{R}}{\partial\Gamma} \left(\frac{E_o}{g_{do}}|\beta=0 \right) \bigg] d\alpha_0 \,d\beta_0.
\label{eq:final_reflection_patch}
\end{split}
\end{equation}
Here, we have left some dependencies implicit for brevity, such as the $r$ and $\phi$ dependence of $g_{do}$ and $\tau$, and the dependence of $\mathcal{R}$ on the cosine of the emission angle, $\mu_e$.
Integrating Eq.~\ref{eq:final_reflection_patch} over all impact parameters corresponding to geodesic paths that intersect the accretion disc, we calculate the total reflection spectrum seen by the distant observer. 

\subsection{Total Emission}
The total emission crossing the distant observer is simply the sum of the direct emission from the source and the re-processed emission from the disc. 
To use the transfer function formalism, we express the emission as a sum of convolutions. The total specific flux is 
\begin{equation}
\begin{split}
S(E_o,t) =  &\,\, D(E_o) \left[A(t') + B(t')\ln(E_o/g_{so}))\right] + \\
&A(t') \otimes  w_{0}(E_o,t'')- B(t') \otimes  \left[ w_{1}(E_o,t'') + w_{2}(E_o,t'') \right] \\
\end{split}
\label{eq:total_specific_flux}
\end{equation}
where $w_0$, $w_1$ and $w_2$ are the three response functions associated with the three terms on the right hand side of 
Eq.~\ref{eq:final_reflection_patch} (see Appendix~\ref{app:transfer_functions} for the explicit expressions)
and $\otimes$ denotes a convolution.

In the Fourier domain convolutions correspond to multiplications. The Fourier transform of the total observed specific flux is therefore
\begin{equation}
\begin{split}
S&(E_o,\nu)= A(\nu) \bigg[D(E_o)+ W_0(E_o,\nu) \bigg] + \\
&B(\nu) \bigg[ D(E_o) \ln\left(\frac{E_o}{g_{so}}  \right) - W_{1}(E_o,\nu) - W_{2}(E_o,\nu) \bigg]
\end{split}
\label{eq:total_specific_flux_fourier}
\end{equation}
where $W_0$, $W_1$ and $W_2$ are the transfer functions (see Appendix~\ref{app:transfer_functions}), which are the Fourier transforms of the response functions.
Following \citet{Mastroserio2018}, we calculate the complex covariance in order to fit not only to the time-average energy spectrum but also to the lags and covariance amplitude as a function of energy for different Fourier frequencies. Since this exploits more information in the signal than time-averaged spectroscopy alone, it enables better constraints of the system parameters. 
The model complex covariance is then $G(E_o,\nu) = S(E_o,\nu) \,e^{- i \phi_r(\nu) }$, where $\phi_r(\nu)$ 
is the phase of the reference band in our model. 
The complex covariance is therefore the cross-spectrum divided through by the modulus of the reference band. 
Following the notation of \citet{Mastroserio2018}, this gives
\begin{equation}
\begin{split}
G(E_o,&\nu) = \alpha(\nu) \bigg[ e^{i\phi_A(\nu)} D(E_o) + \gamma(\nu)e^{i\phi_B(\nu)} \ln\left(\frac{E_o}{g_{so}}\right)D(E_o) + \\
&e^{i\phi_A(\nu)} W_o(E_o,\nu)-e^{i\phi_B(\nu)}W_1(E_o,\nu)-e^{i\phi_B(\nu)}W_2(E_o,\nu) \bigg],
\end{split}
\label{eq:complex_covariance}
\end{equation}
where $\gamma(\nu)$ is the relative amplitude $|B(\nu)|/|A(\nu)|$ and the phase of the reference band, $\phi_r(\nu)$, is swallowed up in the definitions 
of the phases $\phi_A(\nu)$ and $\phi_B(\nu)$, 
which are left as model parameters for a given frequency range, 
as are $\alpha(\nu)$ and $\gamma(\nu)$.
This expression contains the non-linear effects considered in \citet{Mastroserio2018} 
and additionally all GR effects, as in Ingram et al (2019).

\section{Data}
\label{sec:data}

As in \citet{Mastroserio2018}, we consider archival \textit{RXTE} observations of Cygnus X-1 from the proposal number P10238. We stack together the final five of the seven observations in this data set\footnote{Observation IDs: 10238-01-06-00, 10238-01-07-00, 10238-01-07-000, 10238-01-08-00, 10238-01-08-000.}, since their spectra are very similar to one another in shape. We discard the first two observations because their spectra are clearly different to the rest. The total exposure is $56.2$~ks for the time-average energy spectrum and $46.6$~ks for the complex covariance. The difference in exposure time results from data discarded when selecting segments of contiguous data in order to perform ensemble averaging. It is worth mentioning this data set has been extensively used in the past \citep[e.g.][]{Revnivtsev1999,Kotov2001}, and even recently 
\citep[e.g.][]{Mahmoud2018a}, because it has high count rates, excellent timing resolution and adequate energy resolution.

Using only Proportional Counter Array (PCA) data, we extract a $2.84$ - $3.74$ keV reference band light curve and $28$ `subject band' light curves in the energy range $4-20$ keV using the exact procedure described in \citet{Mastroserio2018}. 
As in our previous work \citep{Mastroserio2018}, we calculate the cross-spectra between each subject band light curve and the reference band light curve, employing ensemble averaging as well as averaging over broad frequency ranges. 
We calculate the complex covariance by dividing the averaged cross-spectrum by the square root of the reference band (Poisson noise subtracted) power spectrum. The complex covariance is a complex quantity as a function of energy and frequency. 
Its modulus is related to the variability amplitude of the signal through the coherence \citep{Vaughan1997} which we assume to be close to unity \citep{Nowak1999}. 
We use \textsc{xspec} version 12.10 \citep{Arnaud1996} to fit our model \textsc{reltrans} (Ingram et al 2019), modified from the publicly available version to additionally account for non-linear effects, simultaneously to the real and imaginary parts of the observed complex covariance for $10$ frequency ranges and to the time-averaged spectrum.
We add $0.1\%$ systematics only to the time-average energy spectrum. 
Considering real and imaginary parts of the complex covariance instead of amplitude and phase allows us to extract the same information from the data and has many advantages, such as easily accounting for the telescope response within \textsc{xspec} (\citealt{Mastroserio2018}; Ingram et al 2019).


\section{Fit to the time-averaged Spectrum }
\label{sec:en_spectrum}
\begin{figure}
	\includegraphics[width=\columnwidth]{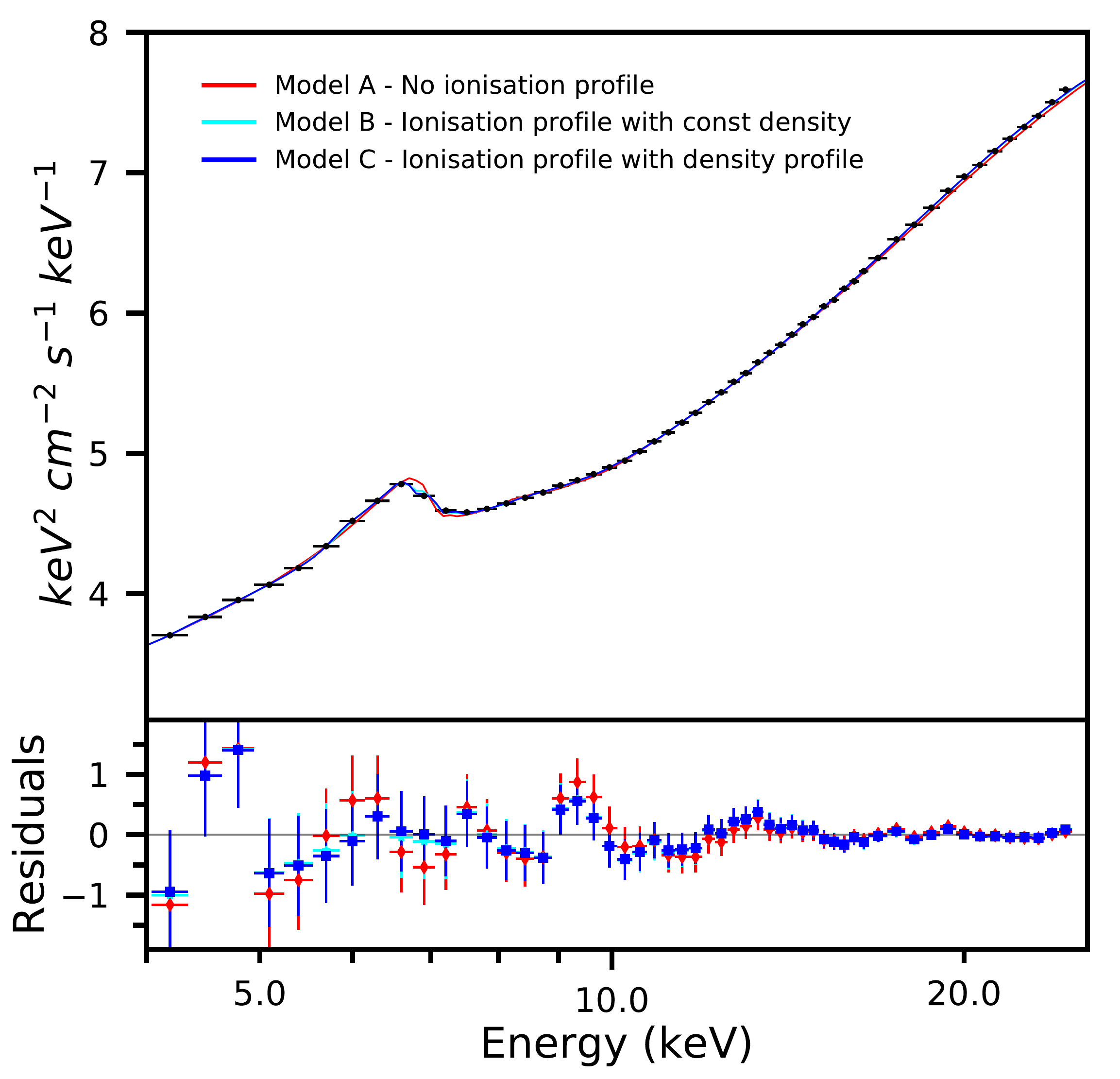}
    \caption{Unfolded spectrum (upper panel) and residuals (lower panel) of the time averaged energy spectrum fit. The three colours represent different models explained in the text (the data points are unfolded around the best fitting model C). $0.1\%$ systematics have been added to the spectrum. A fit with no systematics added produces the same parameter values and residual shape.}
    \label{fig:DC_eeuf}
\end{figure}

We start by fitting our model \textsc{reltrans} to the time-averaged spectrum alone. 
Line-of-sight absorption is accounted for within the \textsc{reltrans} model using the \textsc{xspec} intrinsic model \textsc{TBabs}. 
We assume the abundances of \citet{Wilms2000} throughout our analysis. 
We explore three different assumptions regarding the disc ionisation state within the 
\textsc{reltrans} model, which are described in the following sub-section.

\subsection{Models}

\textsc{reltrans} self-consistently calculates the normalisation of the reflected component of the observed spectrum relative to the direct component assuming a stationary lamppost source. Since in reality the source may be neither stationary nor point-like, the relative normalisation used in the model is equal to the self-consistently calculated value multiplied by the model parameter $1/\mathcal{B}$. Setting this boost parameter $1/\mathcal{B}$ to greater than (less than) unity mimics the effect of plasma moving towards (away) from the disc. The shape of the observed reflection spectrum can additionally be influenced by the spin of the black hole \citep{Martocchia2000}. We note that the model is computationally not very sensitive to the actual spin value \citep[see also][]{Dauser2013} and 
the leading order effect of the spin is to modify the innermost stable orbit (ISCO) of the accretion disc. In order to probe the full extent of the disc we fix the spin parameter to $0.998$ throughout and allow $r_{in}$ to be a free parameter.

The restframe spectrum, calculated using the model \textsc{xillver} \citep{Garcia2010,Garcia2013}, depends on the disc ionization parameter $\xi = 4\pi F_x/n_e$, where $F_x$ is the $13.6$ eV to $13.6$ keV illuminating flux and $n_e$ is the electron density of the disc (which is assumed to be vertically constant for the \textsc{xillver} calculation). This parameter sets the upper boundary condition of the \textsc{xillver} calculation for the vertical temperature and ionization balance in the disc upper atmosphere. In most spectral studies, $\xi$ is assumed to be constant over the entire disc extent. However, our model introduces an ionisation profile as a function of radius (following \citealt{Svoboda2012,Chainakun2016a,Kammoun2019}). 
Numerically, this involves discretizing the calculation into a number of ionization zones. Throughout this paper we use 10 ionization zones, which is found to provide adequate resolution (Ingram et al 2019).

We consider three different ionisation profiles. Model~A uses the same ionisation parameter for the entire disc. This model is identical to the model \textsc{relxilllp} (\citealt{Dauser2013,Garcia2014}; Ingram et al 2019). Model~B uses an ionization profile determined by self-consistently calculating $F_x(r)$ in the lamppost geometry and assuming $n_e=$ constant (as in \citealt{Svoboda2012}). Model~C again uses the same self-consistent calculation of $F_x(r)$, but assumes the following density profile
\begin{equation}
n_e \propto r^{3/2}\left[1 - \left(\frac{r_{in}}{r}\right)^{1/2}\right]^{-2}.
\label{eq:n_e}
\end{equation}
This corresponds to zone A of a \citet[][]{Shakura1973} accretion disc (where the pressure is dominated by radiation and the opacity is dominated by electron scattering; see in particular their Eq. 2.11), and assumes zero torque at the inner disc boundary. The ionisation profile in Model~C is therefore not a monotonic function of radius: it has a maximum located a few $R_g$ outside $R_{in}$ (see Ingram et al. 2019). In all cases, we leave the normalisation of the ionisation profile as a free parameter in the fit.

\begin{table*}
	\centering	
	\caption{Best fitting parameters obtained from fitting to the time-averaged energy spectrum with three different models. The models differ in the ionisation profile: Model~A has a constant ionisation in the entire disc, Model~B has a monotonic ionisation profile peaking at inner radius of the disc and Model~C has a ionisation profile calculated assuming a density profile in the disc given by equation (\ref{eq:n_e}). The spin value is fixed to $0.998$. Errors  are all $90\%$ confidence. Note that while we quote the observed high energy cut-off $(E_{cut,o})$, the cut-off in the source restframe is  $E_{cut}=E_{cut,o}/g_{so}$.}
	\label{tab:DC_par}
	\begin{tabular}{ c c c c c c c c c c c } 
		\hline
		 &$ N_{\rm h} $ $\left(10^{22} \,\,{\rm cm}^{-2}\right)$ & $h \,(R_g)$ & Incl $({\rm deg})$ &$ r_{\rm in} \,(R_g) $ & $\Gamma$ & $\log \xi ^{(a)}$ &$A_{\rm Fe}$ & $E_{\rm cut,o}$ (keV)& $1/\mathcal{B}$&red $\chi^2$\\
		\hline
		A)&$0.7^{+0.1}_{-0.1} $  & $12.3^{+3.4}_{-5.0}$& $30.1^{+1.8}_{-2.1} $ & $ 5.07^{+2.35}_{-^{(b)}}$& $1.70^{+0.01}_{-0.02}$	&$3.02^{+0.02}_{-0.02}$& $1.9^{+0.3}_{-0.2}$& $774^{+156}_{-237}$ & $0.3^{+0.02}_{-0.02}$ & $ 46.67/44$\\		
		\hline
		B)& $0.9^{+?^{(c)}}_{-0.2}$ &$15.5^{+2.6}_{-1.8}$  & $31.1^{+2.9}_{-3.1}$& $3.1^{+0.9}_{-1.2} $ & $1.68^{+0.04}_{-0.04} $& $4.7^{+0.5}_{-0.4}$ &$1.6^{+0.5}_{-0.2}$& $468^{+122}_{-41}$ & $0.45^{+0.02}_{-0.02}$ & $43.52/44$  \\		
		\hline
		C)&$ 0.9^{+?^{(c)}}_{-0.3}$  & $16.5^{+2.4}_{-4.7}$& $32.1^{+2.6}_{-2.5} $ & $4.8^{+0.9}_{-0.5} $& $1.69^{+0.01}_{-0.02}$ &$3.9^{+0.1}_{-0.6}$& $1.52^{+0.05}_{-0.22}$& $431^{+49}_{-49}$ & $0.45^{+0.03}_{-0.02}$ & $41.37/44$ \\		
		\hline
	\end{tabular}
	\begin{list}{}{}
		\item[$^a$] In Model~B and C the parameter is the peak value of the ionisation profile. 
		\item[$^b$] The lower bound is unconstrained.
		\item[$^c$] The uper bound is unconstrained.
        \end{list}
\end{table*}

\subsection{Results}

Figure~\ref{fig:DC_eeuf} shows the all three models and the data points unfolded around the Model~C best fit. 
The residuals, which are defined as the observed counts minus the folded model, shown in the lower panel are slightly larger in the case of Model~A, and the $\chi^2$ value for this model is correspondingly higher (see Table~\ref{tab:DC_par}). Table~\ref{tab:DC_par} lists the best fitting parameter values for the different configurations of the model. We report all the parameter values in this paper with errors at $90\%$ confidence. 

For all three models, we measure an inclination similar to that inferred for the binary system 
\citep{Orosz2011}.
The inner radius is close to the ISCO but it is found not to be pegged to it, 
as was the case for our previous work on Cygnus X-1 \citep{Mastroserio2018}. 
For Model~A, we were not able to constrain a $90\%$ confidence
lower bound, while in the other two models $r_{in}$ is well defined. 
In contrast with \textsc{reltrans} fits to the time-averaged spectrum of GRS~1915+105 (Shreeram \& Ingram 2019), we do not find that including an ionization profile systematically increases the best fitting value of $r_{in}$. 
The ionization value reported in Table~\ref{tab:DC_par} is the \textit{peak} ionization value in the disc for the case of Models B and C, and thus the value presented in the table is much higher for these two models than for Model~A. The relative iron abundance is mildly super-solar for all three models.

All three models have formally acceptable $\chi^2$ values, although it is encouraging that the configurations featuring the more physical assumption of non-constant radial ionization profile (Model~B and C) have the lowest $\chi^2$ values. All three fits are also similar in terms of residual systematics (Fig.~\ref{fig:DC_eeuf}), making it difficult to distinguish between them. 
We note that some parameters are nearly degenerate, for example, the absorption column density $(N_h)$ and 
the high energy cut-off $(E_{cut})$, and the height of the point source $(h)$ and the inner radius $(r_{in})$ are correlated in the fit. 

\section{Reverberation mapping and mass measurement}
\label{sec:rev_mapping}

\begin{figure}
	\includegraphics[width=\columnwidth]{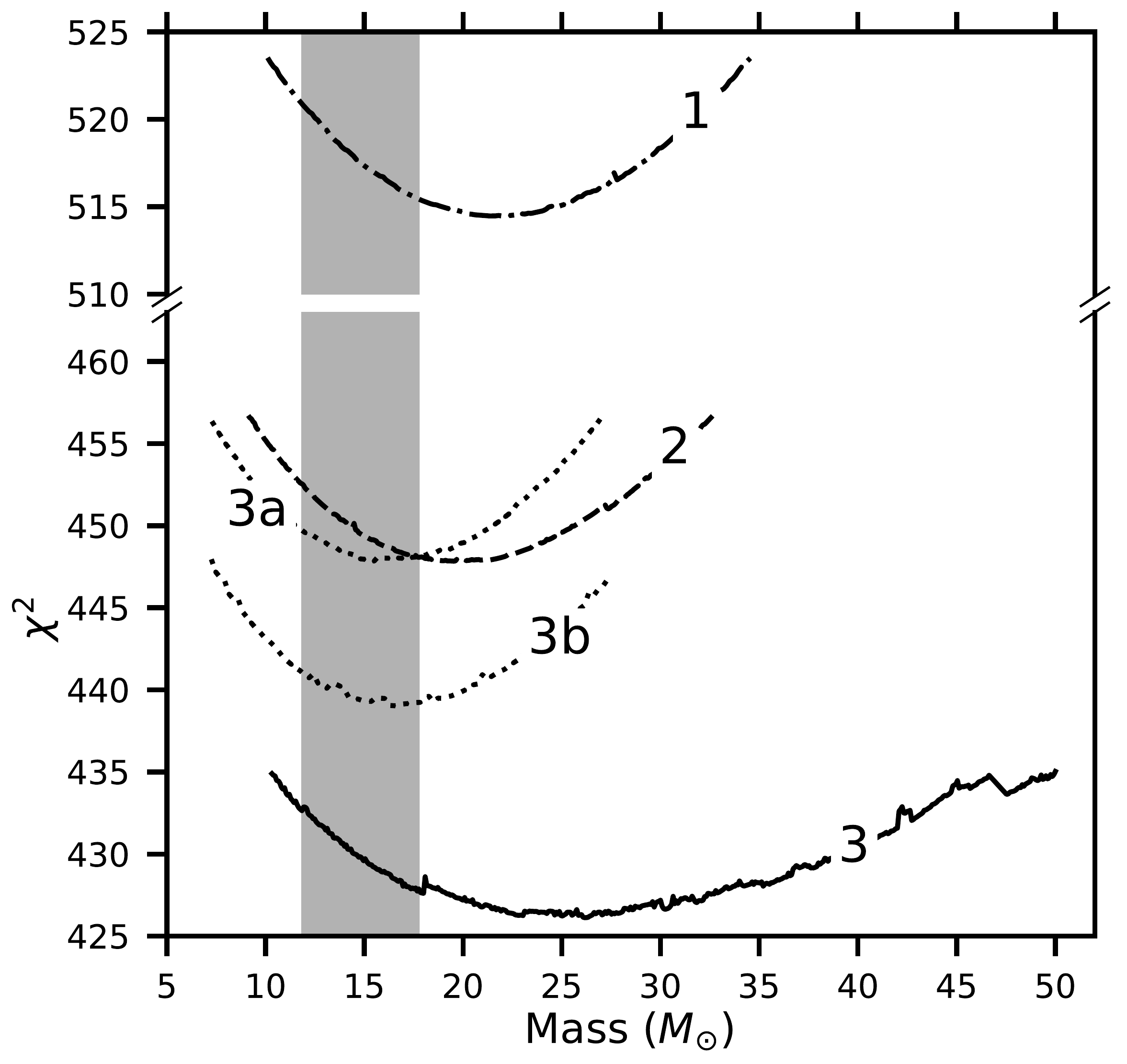}
    \caption{Chi-squared curves of black hole mass calculated for different assumptions on the degree of ionisation of the disc as a function of radius. The shaded region represents the $3\sigma$ confidence interval of the existing dynamical mass measurement. }
    \label{fig:contour_mass}
\end{figure}

We now conduct joint fits, simultaneously considering the real and imaginary parts of the complex covariance across $10$ frequency ranges and the time-averaged spectrum. 
We follow the approach of \citet{Mastroserio2018}, except now we consider a broader range of frequencies ($0.1$ mHz to $32$ Hz) and leave the black hole mass as a free model parameter in the fits. Fig.~\ref{fig:contour_mass} shows the $\chi^2$ curves of the mass for five different configurations of the model (described below). 
It is worth noting that all the curves constrain the mass to be between $5$ and $50$ $M_{\odot}$, which is reasonable for Cygnus X-1 (all the curves stop at $3\sigma$ likelihood).
The shape of the curves demonstrates that the method can constrain the mass of the compact object even with archival \textit{RXTE} data.
The grey area in Fig.~\ref{fig:contour_mass} represents the $3\sigma$ confidence interval of the existing dynamical mass measurement \citep{Orosz2011}. We see that all the curves intersect the grey box. All the configurations are described in the following subsections and the model parameters are listed in Table~\ref{tab:cov_par}. 

\begin{figure}
	\includegraphics[width=\columnwidth]{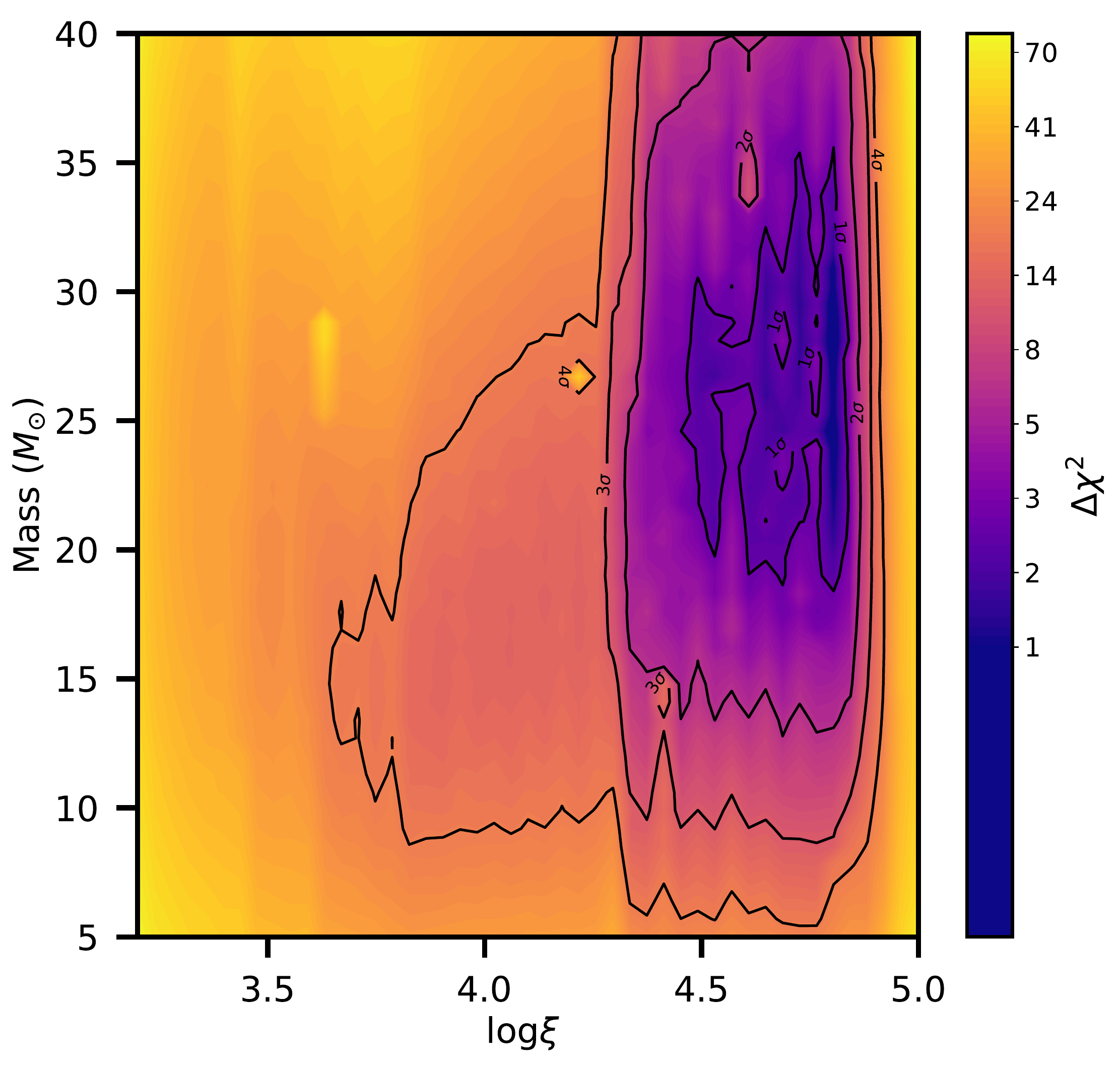}
    \caption{$\Delta \chi^2$ 2D contour plot between the black hole mass and the peak ionisation parameter. The ionisation profile is the one used in  Model~3. The colour scale indicates the relative increase of $\chi^2$ compared with the best fit of Model~3. The 4 black lines are the sigma contours corresponding to 2 degrees of freedom ($\Delta \chi^2=$ $2.3$, $6.18$, $11.83$ and $19.33$, corresponding to $1$, $2$, $3$ and $4$ $\sigma$).}
    \label{fig:2Dcontour}
\end{figure}

\subsection{Model~1: constant ionisation profile}
\label{subsec:model1}
The first configuration has a constant ionisation profile (the same as Model~A in the previous section).
The hydrogen column density has a very low value compared to what was used in other works \citep{Gilfanov2000,Tomsick2014} and its lower limit is not constrained by the fit, which is also the case for the disc inner radius.
Although the mass is formally compatible with the dynamical measurement, it is larger than the expected value.

\subsection{Model~2: ionisation profile calculated assuming constant density}
\label{subsec:model2}
The second configuration calculates $\xi(r)$ assuming constant disc density (the same as Model~B in the previous section), and the peak value of $\xi(r)$ is set as a free model parameter.
In this configuration, $\xi(r)$ always peaks at $r_{in}$, and the ionisation value there indicates that 
the plasma is approximately fully ionised (complete ionisation occurs at $\log_{10}\xi \approx 4.7$). 
However, further out the ionisation decreases, so there is still a weak iron $K\alpha$ 
line feature in the \textsc{xillver} spectrum for $\log_{10}\xi=4.5$,
and the inner radius is still constrained by the overall shape of the iron line.
The $\chi^2$ is lower than for Model~1 and the mass is slightly closer to the dynamical value

\subsection{Model~3: ionisation profile calculated assuming zone A density profile}
\label{subsec:model3}
The third configuration calculates $\xi(r)$ assuming that the electron density depends on radius following equation (\ref{eq:n_e}), which is appropriate for zone A of a Shakura \& Sunyaev disc. 
The peak value of $\xi(r)$ is again a free model parameter, and the best-fitting value is again very high, but this peak is now at $r > r_{in}$ because the disc density becomes very high approaching $r_{in}$ due to the boundary condition in Eq.~\ref{eq:n_e}. This model has the lowest $\chi^2$ value, but the $90\%$ lower bound of the best fitting mass is higher than the $90\%$ upper bound of the dynamical measurement.

\subsection{Model 3a and 3b: ionisation profile calculated assuming zone A density profile, peak ionisation parameter bounded}
\label{subsec:model4}
The very high peak values of the best fitting ionization parameter for the previous two models may be implausible. 
Since we know the distance to Cygnus X-1, we can use the observed continuum flux and a reasonable estimate for the disc electron density in order to place the following upper limit on the ionization parameter (see Appendix~\ref{app:ion_calc} for a derivation)
\begin{align}
  &\xi(r) =\, 5.77\times 10^{3} ~{\rm erg}~{\rm cm}~{\rm s}^{-1}
 \left(\frac{\left[1/\mathcal{B}\right]}{0.28}\right) \left(\frac{\epsilon(r) g_{sd}^{2-\Gamma}(r)g^{\Gamma-2}_{so}}{0.0014}\right)   \nonumber \\
& \left(\frac{\mathcal{F}}{1.7\times 10^3~{keV}{cm}^{-2}{\rm s}^{-1}}\right) \left(\frac{D}{2.5{\rm kpc}}\right)^2 \left(\frac{4\times 10^{20}{\rm cm}^{-3}}{n_e}\right) \left(\frac{10 M_{\odot}}{M}\right)^2 .
\label{eq:ion_calc} 
\end{align}
Here, $\mathcal{F}$ is the integral from $13.6$ eV to $13.6$ keV of the source flux (in the source rest frame) in our model (see Appendix~\ref{app:ion_calc} for the exact definition).
The values quoted in brackets are chosen as reasonable limits of each term that maximize $\xi(r)$. The first three terms on the right hand side of Eq.~\ref{eq:ion_calc} can be derived directly from our fits. The values we use for these terms are derived from the Model~1 fit (which maximizes the estimate of $\xi(r)$, mostly due to the low source height).
We estimate an upper limit for the distance to Cygnus X-1 from the \textit{Gaia} second data release ($2.17\pm 0.12$ kpc; \citealt{Gandhi2019}). This is higher than the radio parallax distance of $1.86^{+0.12}_{-0.11}$ kpc \citep{Reid2011}. We choose the \textit{Gaia} distance in order to get an upper limit.
We consider a conservative lower limit for the mass of $10 M_{\odot}$. 
For our estimated lower limit of the electron density, we use the value of $\sim 4\times 10^{20} cm^{-3}$ measured by \citet[][]{Tomsick2018}, which is lower than analytic estimates from disc theory \citep{Svensson1994,Garcia2016}.
We can therefore derive from Eq.~\ref{eq:ion_calc} the highest physical upper limit on the peak ionization parameter $\log_{10}\xi_{\rm max} $ of $\approx 3.76$, implying that the best fitting values of this parameter for Models 2 and 3 are implausibly high.
We therefore define Model~3a, which applies the same assumptions as Model~3 except now we set a sensible upper limit on $\log\xi_{\rm max}$ of $3.5$. 
Although the minimum $\chi^2$ for this new model is higher than for Model~3 (Fig.~\ref{fig:contour_mass}), the new model is more physical. It returns a mass value closer to the existing dynamical measurement. Finally, we define Model~3b, in which we relax the upper limit on the peak ionization parameter to $\log\xi_{\rm max} \leq 4$. As expected, the $\chi^2$ value is lower than for Model~3a and the best fitting mass increases slightly.

Fig.~\ref{fig:2Dcontour} shows the correlation between black hole mass and peak ionisation parameter in more detail in the form of a 2D $\chi^2$ contour plot for Model 3. In line with Fig.~\ref{fig:contour_mass}, we see that the minimum $\chi^2$ value corresponds to a very high ionisation and a black hole mass of $\sim 25 M_{\odot}$. For lower (more realistic) values of peak ionisation, a lower mass ($\sim 15 M_{\odot}$) is preferred. Very low peak ionisation values are strongly ruled out.

\subsection{Other Model Parameters}
\label{subsec:par_models}

\begin{table*}
	\centering	
	\caption{Best fitting parameters obtained from our simultaneous fit to the complex 
		covariance in $10$ frequency ranges ($0.98mHz - 32 Hz$) and the time-average spectrum. 
        The spin value is fixed to $0.998$.	The models listed in the table have different ionisation profiles. 
        Model~1 has a constant ionisation profile and $\xi$ is the ionisation value throughout the disc.
        Model~2 has an ionisation profile calculated assuming constant density and self-consistently calculated illuminating flux.
        Model~3 has an ionisation profile calculated assuming a `Zone A' density profile and self-consistently calculated illuminating flux.
        Models 3a and 3b have the same ionisation profile as Model~3, except the peak ionization is pegged to an upper limit of $3.5$ and $4.0$ respectively. Errors  are all $90\%$ confidence.
        Note that while we quote the observed high energy cut-off $(E_{cut,o})$, the cut-off in the source restframe is  $E_{cut}=E_{cut,obs}/g_{so}$.}
	
	\begin{tabular}{ c c c c c c c c c c c c } 
		\hline
		 &$ N_{\rm h} $ $\left(10^{22} {\rm cm}^{-2}\right)$ & h $(R_g)$  & Incl $({\rm deg})$ &$ r_{\rm in} \, (R_g)$ & $\Gamma$ & $\log \xi ^{(a)}$ &$A_{\rm Fe}$ & $E_{\rm cut,o}$ (keV)& Boost& Mass&red $\chi^2$\\
		\hline
		1)&$0.13^{+0.08}_{-?^{(b)}}$ &$6.2^{+0.6}_{-0.8}$&$32.4^{+0.6}_{-1.0}$&$1.8^{+0.4}_{-^{(c)}}$&$1.60^{+0.01}_{-0.01}$&$3.08^{+0.02}_{-0.01}$&$3.7^{+0.1}_{-0.1}$&$218^{+13}_{-9}$&$0.28^{+0.01}_{-0.01}$&$21.6^{+6.8}_{-6.6}$&$515/563$\\
        \hline
		2)&$0.7^{+0.2}_{-0.1}$  & $9.7^{+1.9}_{-1.4}$& $33.8^{+1.4}_{-1.5} $ & $ 2.2^{+0.9}_{-1.0} $& $1.65^{+0.02}_{-0.02}$	&$ 4.5^{+0.6}_{-0.7}$&$2.0^{+0.4}_{-0.2}$&$371^{+55}_{-63}$ &$0.50^{+0.05}_{-0.04}$& $19.7^{+5.3}_{-5.4}$&$ 448/563 $\\		
		\hline
		3)&$0.6^{+0.1}_{-0.1}$ & $9.0^{+1.4}_{-1.7}$ & $35.0^{+1.2}_{-1.2}$ & $1.3^{+0.7}_{-^{(c)}}$ & $1.67^{+0.01}_{-0.01}$ & $4.7^{+0.7}_{-0.6}$ & $1.7^{+0.2}_{-0.3}$ & $712^{+82}_{-97}$ & $1.1^{+0.2}_{-0.1}$ & $ 26.0^{+9.6}_{-8.6}$&$426/563$\\
		\hline

        3a)&$0.79^{+0.04}_{-0.10}$ & $9.5^{+2.7}_{-3.9}$ & $35.1^{+1.4}_{-1.0}$ & $5.9^{+0.7}_{-0.7}$&$1.67^{+0.01}_{-0.02}$& $ 3.5^{(d)}$ &$1.8^{+0.2}_{-0.1}$&$377^{+36}_{-36}$&$0.46^{+0.02}_{-0.03}$&$16.5^{+5.0}_{-5.0}$
        &$447/563$\\        
		\hline
        3b)&$0.75^{+0.09}_{-0.24}$ & $10.8^{+1.9}_{-1.2}$ & $35.0^{+0.8}_{-1.8}$ & $3.1^{+0.4}_{-0.4}$&$1.66^{+0.02}_{-0.02}$& $ 4.0^{(d)}$ &$1.7^{+0.5}_{-0.3}$&$355^{+54}_{-96}$&$0.60^{+0.04}_{-0.03}$&$16.5^{+6.1}_{-5.2}$
        &$439	/563$\\        
        
		\hline

\end{tabular}
	\begin{list}{}{}
		\item[$^a$] Besides Model~1 the parameter is the peak value of the ionisation profile. 
 		\item[$^b$] The lower limit of the hydrogen column density is not constrained below $0.2$
 		\item[$^c$] The lower limit of the inner radius is the ISCO $(1.237R_g)$.
   		\item[$^d$] The ionisation peak is pegged to the upper limit.
	\end{list}
    \label{tab:cov_par}
\end{table*}
Table~\ref{tab:cov_par} lists best fit parameter values for the different configurations of the model.
For all models except Model~1, the hydrogen column density agrees with the value of $N_h\approx 0.6\times10^{22} \, {\rm cm}^{-2}$ previously used for this data set \citep{Gilfanov2000,Grinberg2014,Tomsick2018}. 
Model~1 again differs from the others with regard to the iron abundance, which is twice solar or less for all models with an ionization profile. Previous time-averaged spectral fitting studies have returned much higher iron abundances for Cygnus X-1 (up to $\sim 5$ times solar \citealt{Parker2015,Walton2016}) and other sources \citep[e.g. GX 339-4][]{Garcia2015,Parker2016}, as did our previous spectral-timing analysis that included neither light bending nor a radial ionization profile \citep{Mastroserio2018}.
Our fits to the complex covariance all return slightly higher values of disc inclination angle than our fits 
to the time-averaged spectrum alone (see Table~\ref{tab:DC_par}), which are in turn all larger 
than the inclination angle of the binary system \citep{Orosz2011}. 
However, our fits return lower inclinations than some previous time-averaged spectral fitting studies in both the soft
\citep[> 38\degree][]{Walton2013,Tomsick2014} and hard 
\citep[> 42\degree][]{Parker2015} spectral states. 
This discrepancy is not down to differences in the fitted models, since our single ionisation model for the time-averaged spectrum is identical to \textsc{relxilllp}, which was used for the other studies, yet our Model A fit to the time-averaged spectrum returns a low inclination. The previous studies used different data sets, some from other instruments. Interestingly, we find that fitting for the complex covariance \textit{without} also considering the time-averaged spectrum returns still higher inclination values (between 39\degree and 45\degree).

When expressed in units of $R_g$, the disc inner radius $r_{in}$ is very small for the first three configurations, 
and becomes larger for the final two models with limits on the peak ionization. 
We find that the same happens when we place an upper limit on the peak ionization parameter for Model~2. 

There is clearly an anti-correlation between the ionisation parameter and the inner radius of the disc that can be seen by comparing different fits and confirmed by the error contour plot (see Appendix\ref{app:MCMC}).
The data can not accommodate a strong contribution from
the innermost part of the disk. 
Either this innermost part is mostly ionised, or,
if the ionisation is forced to be lower,
the data requires a slightly truncated disc to exclude the 
contribution from the innermost region of the disc. 

The source height for all our models is significantly larger than the very small value ($h \lesssim 2~R_g$) reported by \citet{Parker2015}. \citet{Niedzwiecki2016} pointed out that such a small source height implies an intrinsic source luminosity many times the Eddington limit due to the large gravitational redshift experienced by source photons. 
Our models do not suffer from such a problem, with the intrinsic source flux only being a factor $\sim 1.33$ times the observed source flux.
We note that our previous analysis in which we ignored light bending and fixed the mass to $14.8 M_{\odot}$ instead needed $h< 2.5\, R_g$ \citep{Mastroserio2018}. The larger source height returned by our current analysis is not down to the mass being a free parameter, since a larger black hole mass would require a \textit{lower} source height to reproduce the same reverberation lag. Instead, the higher source height is well explained by the inclusion of light bending, since this leads to an increased reverberation lag for a given geometry. We therefore now need a higher source height to explain the same data.
The source heights we measure here are similar to those found by e.g. \citet{Basak2017} and \citet{Tomsick2018}, although direct comparison is difficult since these studies employed different model assumptions.

Introducing an ionization profile seems to increase the high energy cut-off returned by the fit, particularly for Model~3.
The values returned here are higher than for previous studies (e.g. \citealt{Wilms2006,Mastroserio2018}).

In order to further explore correlations between parameters, we run a Markov Chain Monte Carlo (MCMC) parameter exploration on Model 3. The setup and results of this analysis are presented in Appendix \ref{app:MCMC}.

\subsection{Model comparison}

Of the models we tested, Model~3 has the lowest $\chi^2$ value. We note that some of the parameter values for this model differ from the other models. In particular the boost parameter is close to unity for this model, consistent with a steady point source, but $\lesssim 0.5$ for the other models, consistent with the coronal plasma moving away from the black hole. Moreover both the high energy cut-off and the mass are much higher for Model~3. Interestingly if we fix the boost parameter to $0.5$ in Model~3, the best fit requires an energy cut-off of $\sim 356$ keV, a mass of $\sim 15 M_{\odot}$, the inner radius increases to $\sim 3.7 R_{g}$ and the ionization peak decreases to $\sim 3.8$\footnote{When the boost is fixed to $0.5$ the parameters not mentioned in the text are similar to the ones of the other configurations.}. 
It seems that forcing the boost parameter to $0.5$ aligns Model~3 to Model~3a and 3b, with a similar reduced $\chi^2$ of $ 440/564$.

\begin{figure*}
	\includegraphics[width=\textwidth]{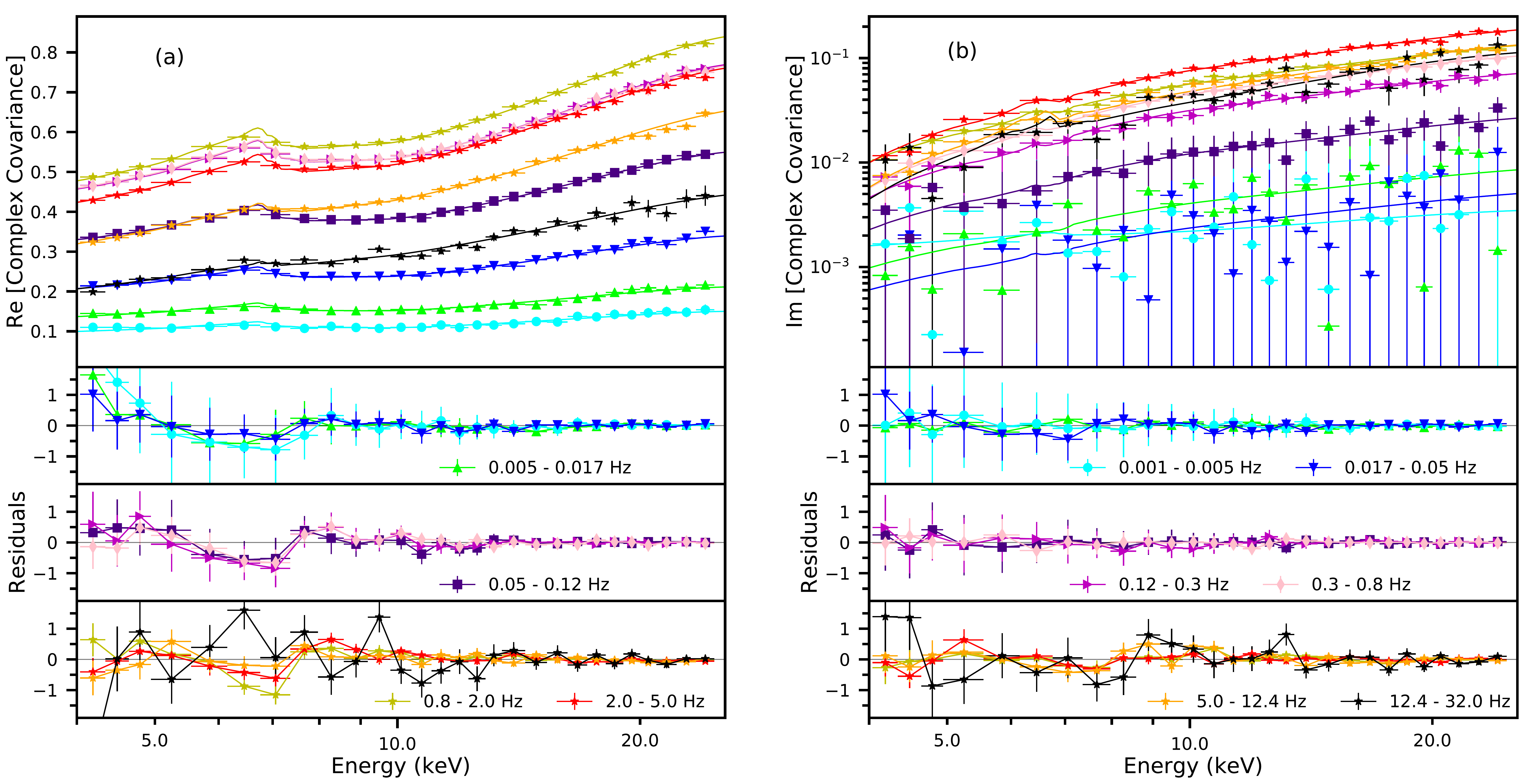}
    \caption{Top panels show the fit of real (a) and imaginary (b) part of the Cygnus X-1 complex covariance spectrum with Model~3. 
    The fit also includes the time-averaged energy spectrum which is not shown in this plot. 
    The dots are the data and the lines in the top panels are the model which has a much higher energy resolution than the data for clarity. 
    The bottom panels show the data minus the folded model 
    (command \texttt{plot rediduals} in \textsc{xspec}).}
    \label{fig:eeuf_res}
\end{figure*}

\begin{figure}
	\includegraphics[width=0.42\textwidth]{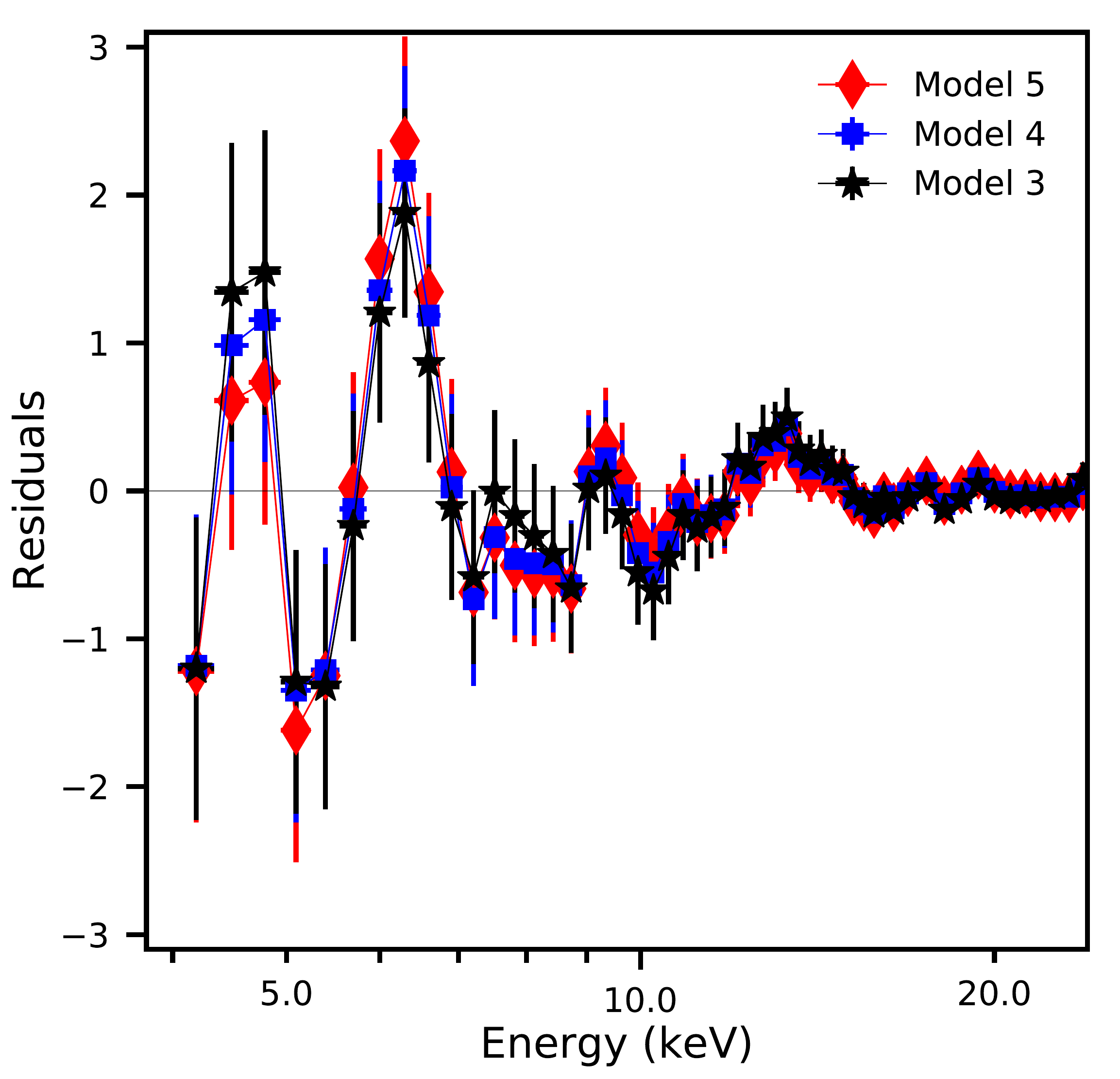}
    \centering
    \caption{Residuals (data minus the folded model) 
    of the time-average energy spectrum of Model~3 (black stars), Model~3a (blue squares) and Model~3b (red diamonds). Here the residual plot shows only the time-averaged energy spectrum contribution for clarity, however this a joint fit with the complex covariance.}
    \label{fig:res_time-ave}
\end{figure}

\begin{figure*}
	\includegraphics[width=\textwidth]{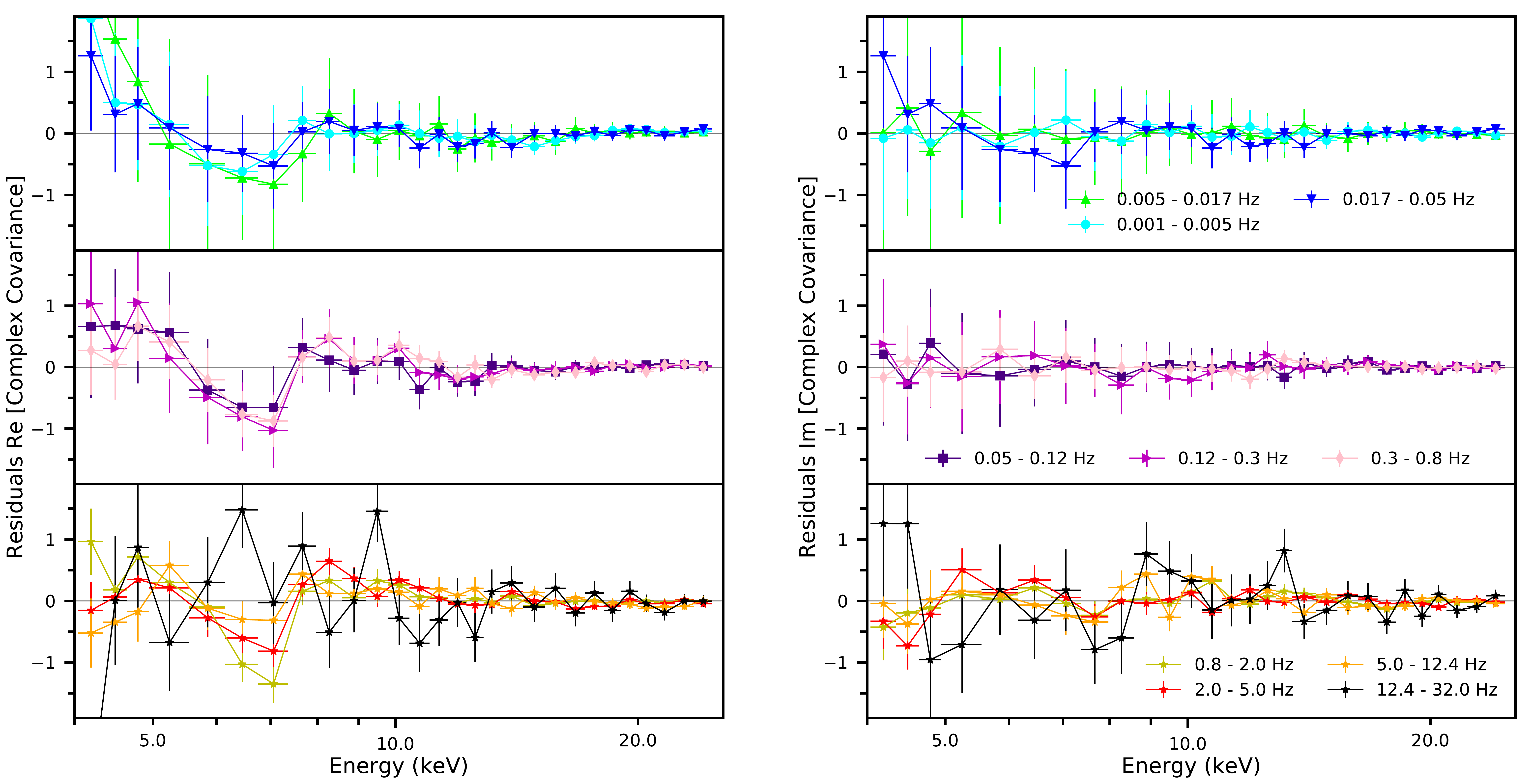}
    \caption{Residuals (data minus the folded model in units of normalised counts per second per keV) of the complex covariance spectra (real and imaginary part) for Model~3a. The fit also includes the time-average energy spectrum which is not shown in this plot.}
    \label{fig:res_M4}
\end{figure*}

Fig.~\ref{fig:eeuf_res} shows the real (a) and imaginary (b) parts of the complex covariance in the frequency range $0.1$ mHz to $32$ Hz. The lines are Model~3 calculated with the best fitting parameters and the points are the data unfolded around the best fitting model. The lower panels in both plots show the residuals of the fit. 
Our current analysis considers three additional low frequency ranges compared with \citet{Mastroserio2018}, which improves the overall signal to noise. However, we do not expect this to have changed the best fitting parameters, since the reverberation time lag does not change its value significantly below the lowest frequency considered in \citet{Mastroserio2018} (e.g. see \citealt{Emmanoulopoulos2014}).
We note that around the iron K$\alpha$ line energy range 
the model slightly overestimates the width of the line.
Similar residuals were found in \citet{Mastroserio2018}, although implementation of light bending and an ionisation 
profile has made the new residuals around the iron line less prominent than in the previous model. 
For the purposes of clarity, this figure does not show the time-averaged spectrum, whose residuals are 
shown instead in Fig.~\ref{fig:res_time-ave}. The black stars are the energy spectrum residuals of 
Model 3 and they show a different structure compared with Fig.~\ref{fig:DC_eeuf}, which represents 
the fit only to the time-average spectrum. It seems that adding the complex covariance to our 
analysis requires the iron line to be broader. This is also indicated by the best fitting value 
for the disc inner radius, which becomes smaller when we include the complex covariance 
in the fit (compare Models A, B and C with Models 1, 2 and 3).
Fig.~\ref{fig:res_time-ave} also includes the energy spectrum residuals for Model~3a 
(blue squares) and Model~3b (red diamonds). These residuals are again very similar to Model~3. 
Fig.~\ref{fig:res_M4} shows the complex covariance residuals for Model~3a, which are similar 
to the Model~3 residuals (see Fig.~\ref{fig:eeuf_res}). Model~3b has an almost identical 
complex covariance residual plot, which is not shown here. Therefore, although setting 
an upper limit for the peak ionization parameter increases the overall $\chi^2$, 
it does not introduce any new features into the residuals.

\begin{figure*}
	\includegraphics[width=\textwidth]{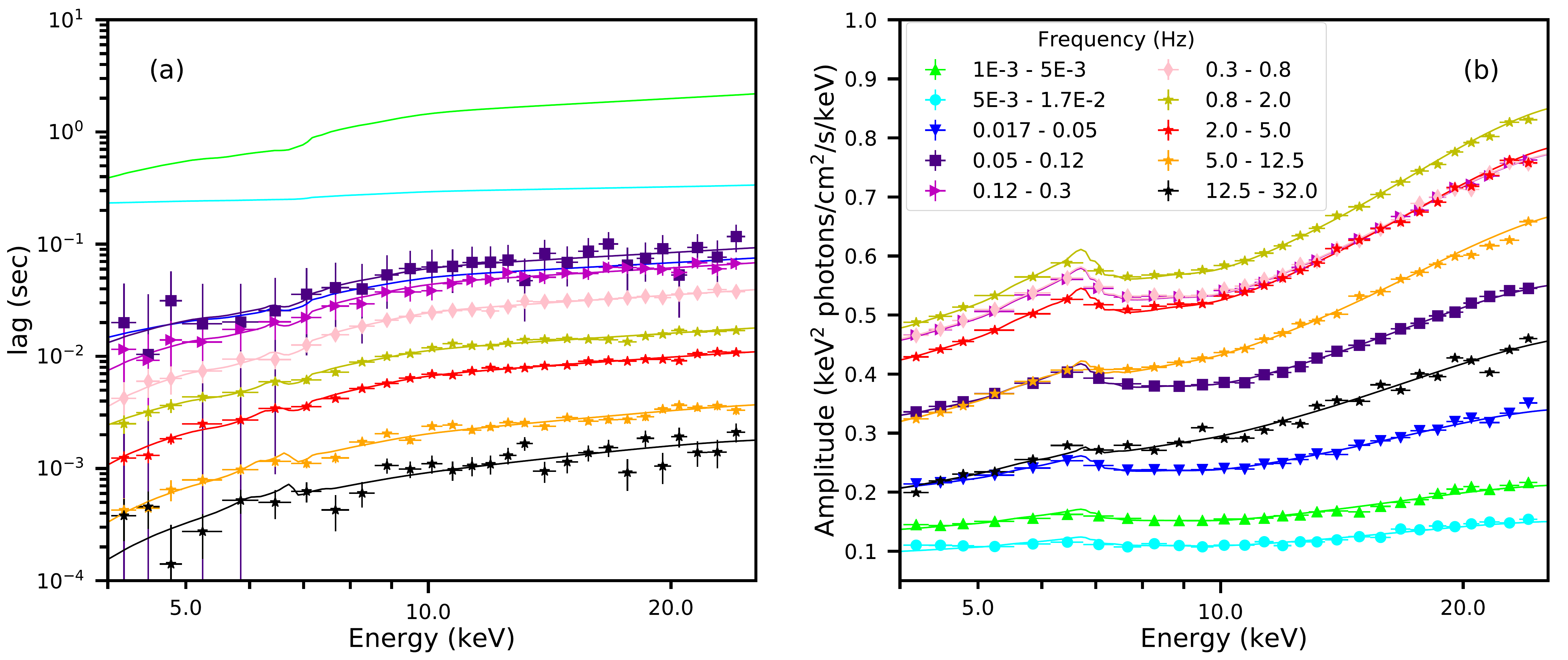}
    \caption{Lags (a) and variability amplitude (b) as a function of energy for different Fourier frequency ranges. 
    The dots are the data and the lines are calculated from the best fitting parameters of Model~3. The data points for the lowest three frequency ranges are not plotted in the lag energy spectrum because their very large error bars. Only the model lines are plotted for these frequency ranges.}
    \label{fig:lag_amp}
\end{figure*}

\begin{figure}
	\includegraphics[width=\columnwidth]{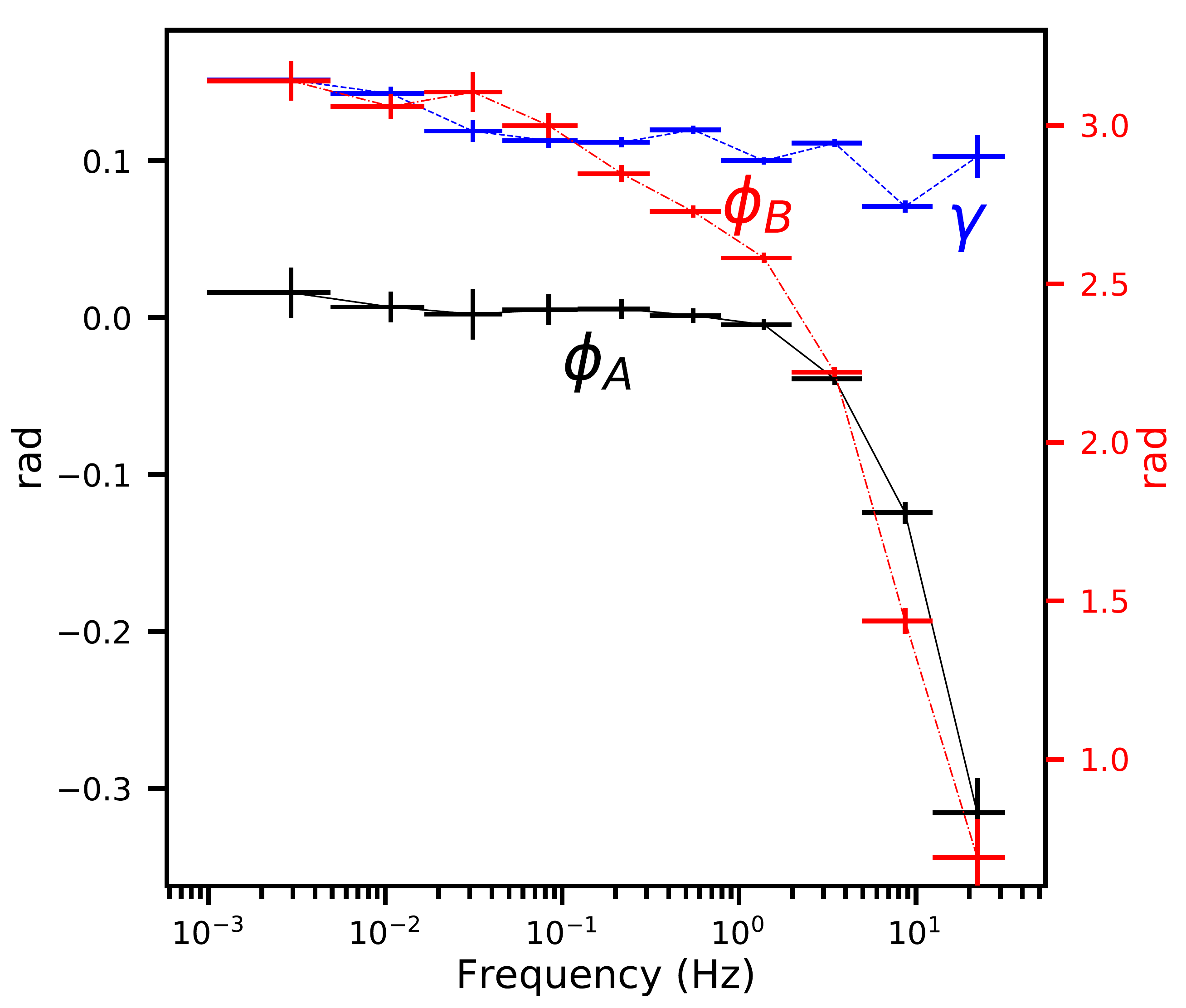}
    \caption{Best fit parameters which control the spectral power-law slope variability, as function of Fourier frequency for Model~3. The $\phi_A$ line and the $\gamma$ line (black solid and blue dashed) refer to the left y-axis. The $\phi_B$ line (red dotted) refers to the right yaxis. Every group of three parameters in the same frequency range has been used to fit the complex covariance.}
    \label{fig:pl_params}
\end{figure}

Although we fit for real and imaginary parts of the complex covariance in order to simply account for the telescope response, we can also plot the time lags and variability amplitude associated with our model, which is more intuitive. Fig.\ref{fig:lag_amp} shows the lags and amplitude for Model~3, with the data points derived from the unfolded complex covariance. For clarity, the model curves are shown with a higher resolution than the data, and the data and the model lags for the frequency ranges $1-5 $ mHz, $5 - 17$ mHz and $17-50$ mHz are not plotted due to their large error bars.
Fig.~\ref{fig:pl_params} shows the continuum parameters as a function of Fourier frequency for Model~3. The behaviour of all three parameters resembles that reported in \citet{Mastroserio2018}, except for $\phi_A$ dropping to slightly lower values than before at high frequency. 
These parameters set the power-law pivoting variations. They are defined in the Fourier domain as the relative phase difference between the reference band light curve and each narrow energy band light curve ($\phi_A, \phi_B$), and as the ratio between the power-law index amplitude and normalisation amplitude ($\gamma$). Although their physical meaning is not immediately clear, it can still be useful to compare them with more physical models.

\section{Discussion}
\label{sec:discussion}

We have, for the first time, estimated the mass of a Galactic black hole through X-ray reverberation mapping. 
We use the model \textsc{reltrans} (Ingram et al 2019) which uses a lamppost geometry 
in full GR to predict complex covariances, modified to include the non-linear effects 
resulting from fluctuations in the slope of the irradiating spectrum. 
The source is located on the black hole spin axis irradiating a razor thin accretion disc in the black hole equatorial plane.

We test a number of models, each employing a different assumption regarding the radial dependence of the disc ionization parameter $\xi(r)$. Model~1 assumes constant $\xi$ and yields a mass of $M=21.6^{+6.8}_{-6.6}~M_\odot$ (errors are $90\%$ confidence). 
Model~2 self-consistently calculates $\xi(r)$ from the radial dependence of the illuminating flux and additionally assuming the electron density in the disc to be constant with radius, yielding a mass of $M=19.7^{+5.3}_{-5.4}~M_\odot$. 
Model~3 self-consistently calculates $\xi(r)$ assuming the radial density profile in zone~A of the \citet{Shakura1973} accretion disc model, and yields a mass of $M=26.0^{+9.6}_{-8.6}~M_\odot$. 
Zone~A (radiation dominated pressure and electron scattering dominated opacity), for a source with luminosity of 1.6\% the Eddington limit, $\alpha = 0.01$ and mass of $14.8\,M_{\odot}$ extends to around $15\, R_g$.
Since the region inside of this dominates the reflection emissivity, the assumption of a zone~A density profile is reasonable. 
Moreover, for radii larger than this, the ionization is low so that its assumed profile is less important.
Model~3 has the lowest $\chi^2$, but we find that the best fitting peak value of $\xi(r)$ is implausibly high. By combining the known distance to Cygnus X-1 with our model parameters and a reasonable estimate for the electron density in the disc atmosphere, we place the highest possible physical upper limit on the peak ionization parameter  $\log_{10}\xi_{\rm max} $ of $ 3.76$.
We therefore define Model~3a, which employs the same assumptions as Model~3 but the peak of the ionization parameter is now bound by the reasonable upper limit on $\log_{10}\xi_{\rm max}$ of $3.5$. 
This is our most physical model and yields a mass of $M=16.5\pm 5~M_\odot$, which is consistent with the existing dynamical mass measurement ($M=14.8\pm1~M_\odot$ \citealt{Reid2011}).
The correlation between the mass and peak ionisation is illustrated clearly by the 2D $\chi^2$ contour plot in Fig~\ref{fig:2Dcontour}.

We note that adding the complex covariance to the fit creates structures
in the residuals of the time-average energy spectrum (Fig.~\ref{fig:res_time-ave}) 
that are not present in the fit to the time-averaged spectrum alone (Fig.~\ref{fig:DC_eeuf}). 
When we fit Model~2 to the complex covariance only (i.e. ignoring the 
time-averaged spectrum), the best fit ionisation peak and inner disc radius 
are compatible with the joint fit, but the mass is larger ($31^{+13}_{-12}\, M_{\odot}$). 
For Model~3, the same experiment instead yields the same mass of the joint 
fit with larger error bars ($M = 25.6^{+13}_{-10}~M_\odot$), but lower 
ionisation peak ($4.0^{+0.3}_{-0.1}$) and slightly larger disc inner radius 
($2.6^{+1.1}_{-1.0}$) (all other parameters are similar to the joint fit).
This, in addition to the Model~3a and 3b fit results, implies an anti-correlation between the inner radius and peak ionisation that is confirmed by an MCMC simulation (see Appendix~C). This anti-correlation occurs because the inner disc does not contribute to the iron line for the highest peak ionization values. Such over-ionization therefore has a similar effect to disc truncation.

Our analysis is sensitive to black hole mass because reverberation lags give distances in units of km and spectral distortions give distances in units of gravitational radii. 
This means that the width of the iron line feature in the time-averaged spectrum \textit{and} in the complex covariance jointly sets geometrical parameters such as the disc inner radius and the source height. Reverberation lags give rise to an iron line feature in the lag-energy spectrum, but this feature is not well constrained for the data set we explore here (see Fig.~\ref{fig:lag_amp} left). 
Still, geometries that would give rise to very large or very small reverberation lags are ruled out, since they would result in lag spectra with respectively a large excess or dip in their lag spectra. 
The energy and frequency dependent correlated variability amplitude on the other hand is very constraining. 
We can clearly see in Fig.~\ref{fig:lag_amp} (right) that the reflection features become weaker for higher Fourier frequencies. This is because the fastest continuum variability is washed out in the reflected emission by the finite size of the reflector (i.e. destructive interference between rays reflecting from different parts of the disc; \citealt{Revnivtsev1999,Gilfanov2000}). 
The larger the path length differences, the more steeply the amplitude of the reflection features will drop off with Fourier frequency. This can be achieved in the model either by increasing the black hole mass, which would not affect the width of the line, or by increasing geometrical parameters such as $h$ or $r_{\rm in}$, which would reduce the width of the line. The model therefore depends on black hole mass in a very different way to any of the geometrical parameters.

\subsection{Future modelling improvements}

Although it is encouraging that our most physical model returns a sensible black hole mass estimate, there are still many improvements that can be made to our reverberation model in future. 
We assume that the disc is infinitely thin whereas in reality the disc has a finite thickness, particularly in the zone~A regime \citep{Shakura1973}. 
Including a realistic disc scale height leads to a more centrally peaked emissivity profile \citep{Taylor2018a,Taylor2018b} leading to shorter time lags and a broader iron line for a given disc inner radius. 
We may therefore expect the future inclusion of a realistic disc height to push the best fitting disc inner radius to larger values, although it is difficult to confidently predict since the new emissivity profile will also effect the ionization profile. 
It has also been suggested that Cygnus X-1's accretion disc is warped, since some reflection spectroscopy results \citep{Tomsick2014,Walton2016} suggest that the inner disc has a different inclination angle to the binary system \citep{Orosz2011}. 
Such a warp would again change the reflection emissivity profile if included in the model. 
However, in other suggestions Cygnus X-1 is an aligned system since it has no peculiar velocity with respect to its association of O stars, perhaps indicating that there was no natal supernova kick to mis-align the black hole spin axis at birth (and maybe even that the black hole was formed via direct gravitational collapse; e.g. \citealt{Mirabel2003}). 
This would also explain the absence of strong low frequency quasi-periodic oscillations (QPOs) in the source in the context of the Lense-Thirring precession model  \citep[e.g.][]{Ingram2009,Rapisarda2017b}. 
We note that our fits here return inner disc inclination angles only slightly discrepant with the binary inclination ($\sim 35\degree$ vs $\sim 27\degree$).

Also, although \textsc{xillver} is state-of-the-art, it will be possible in future to include more realistic disc physics in the calculation of the rest frame reflection spectrum. 
We currently calculate the disc ionization parameter using a reasonable function for the electron density, 
$n_e(r)$, but we use the publicly available \textsc{xillver} grid in order to  
calculate the restframe spectrum, which assumes  $n_e=10^{15}~\rm {cm}^{-3}$. 
We therefore use the correct value of the ionisation parameter for our calculation, 
but the radiative processes and atomic physics are always calculated in a low density environment,
whereas X-ray binary discs are thought to have much higher densities  ($n_e \sim 10^{20} {\rm cm}^{-3}$).
Fixing this issue is challenging, since the atomic data used for the \textsc{xillver} 
calculation is not currently tabulated for the high densities of Galactic
black hole discs (\citealt{Shakura1973,Svensson1994,Tomsick2018}).
The leading order effect of increasing $n_e$ is on the $< 3$ keV spectrum, 
but the importance of using a realistic value has recently been demonstrated 
\citep{Garcia2016,Tomsick2018,Jiang2019}. 
Finally, we assume a point-like stationary corona. 
Alternatively including an extended corona may change our results, 
although it is difficult to judge in which way without performing the necessary calculations.
Properly including a plasma velocity vector may also change our results by adjusting the radial dependence of the illuminating flux as well as just the normalisation \citep{Dauser2013}. 
Our formalism will enable us in future to test our assumptions about the accretion geometry using the known black hole mass and distance for Cygnus X-1, calibrating our model for future use on other sources.

\subsection{Implications}

Our models return a range of values for the disc inner radius, some compatible with earlier claims of near maximal black hole spin \citep{Fabian2012,Parker2015} and some (including our most physical model, Model~3a) with $r_{\rm in}$ as large as $\sim 6~R_g$. 
Whether or not the disc truncates outside of the ISCO in the hard state is still debated. 
Generally, time-averaged reflection spectroscopy tends to indicate smaller disc 
inner radii than does timing analysis \citep[e.g.][]{Rapisarda2017b}. 
We note that a number of the future improvements to our model discussed in the previous subsection could have the effect of pushing the best fitting $r_{\rm in}$ further out, although the parameter space will be rather complex. 
We also note that including a radial ionisation profile does not systematically push $r_{in}$ to larger values, in contrast with the findings of Shreeram \& Ingram (2019) for GRS 1915+105.

In our model, we account for the hard `continuum' lags that dominate at low Fourier frequencies with a pivoting power-law model, and this provides good fits to the data. 
If the corona is indeed quite compact, then such lags could  well be  
produced by varying cooling and heating of the corona. 
Disc photons cool the corona before the propagating accretion rate fluctuations 
arrive into the corona and heat it (Uttley \& Malzac in prep).
The same type of lags would also be present for the case of an extended corona 
but propagating fluctuations \textit{within} the corona (both radially and vertically, 
in the case of a veritical struncture) would need to be taken into account 
in our reverberation model (e.g. \citealt{Wilkins2016,Chainakun2017}). 

The geometry of our best fitting models is compatible with a compact source within which the plasma is moving away from the black hole, perhaps the base of the jet \citep[e.g.][]{Markoff2005, Kara2019}. 
In all cases, our best fitting source height is sufficiently large so as not to require the intrinsic source flux to be super-Eddington after the application of relativistic corrections. However, our fits do return high values of the high energy cut-off. If the corona is stationary, this implies that it must be very extended ($\gg 100~R_g$) in order to be in ${\rm e}^\pm$ pair equilibrium \citep[e.g.][]{Fabian2015}. However, if it is outflowing, then its radiation as we observe it will be Doppler blue shifted due to the fairly low inclination angle, meaning that the rest frame temperature is lower and thus allowing for a more compact corona. More insight into the true rest frame electron temperature can be gained in future by using a continuum model more sophisticated than an exponentially cut-off power-law and by considering data in the $> 25$ keV energy range, as would be provided by \textit{NuSTAR}. We plan to carry out further calibration of the model on well-studied sources such as Cygnus X-1, which will introduce the prospect of using X-ray reverberation mapping to measure the mass of black holes that have no existing dynamical mass measurement. This will be particularly useful for systems that suffer from heavy extinction, and will cut down on systematic errors in cases where the binary inclination angle is poorly constrained. The extra constraints on the Galactic black hole mass function provided by a new mass measurement technique will enhance studies of the mass gap between black holes and neutron stars, and comparison with the mass function of coalescing black hole systems. We note that our method is also applicable to AGN.

\section{Conclusions}
\label{sec:conclusion}
We performed a mass measurement of Cygnus X-1 through X-ray reverberation analysis. 
Our most physical model (Model~3a) requires a black hole mass of 
$16.5 \pm 5\, M_{\odot}$ consistent with the dynamical measurement.
The model uses an accretion disc ionisation profile limited 
to not exceed the highest ionisation that is physically plausible. 
Removing this upper bound (Model~3), leads to a fit that has a lower $\chi^2$ 
but includes physically implausible peak ionisation values, implying either 
a very low density disc atmosphere or a much greater distance to the system
than is inferred from parallax.
We performed our analysis with \textit{RXTE} data. 
Using telescopes with much higher energy resolution such as \textit{NICER} and \textit{NuSTAR} 
would improve our results in terms of constraining the shape of the 
spectral features and disentangle the degeneracy among some of the model parameters.
We plan to use this model in future on such data sets, both for Galactic black holes and AGNs.

\section*{Acknowledgements}
The authors would like to acknowledge the anonymous referee for the very helpful comments and suggestions.
G.M. acknowledges support from NWO. A.I. acknowledges the Royal Society.






\appendix
\section{Transfer functions}
\label{app:transfer_functions}
In the text we derived the expression of the total observed specific flux  as a function of time (equation \ref{eq:total_specific_flux_fourier}). Here we write down explicitly the expressions of the three response functions and their Fourier transform.  
 
\begin{equation}
\begin{split}
    w_0(E_o, t) = \int_{\alpha_0} \int_{\beta_0} &g_{do}^3 \epsilon(r)\, \delta(t-\tau)\\ &\mathcal{R}\left(E_o|\mu_e,\Gamma,g_{sd}E_{cut} \right) d\alpha_0 \,d\beta_0 ,
\end{split}
\end{equation}

\begin{equation}
\begin{split}
    w_1(E_o, t) = \int_{\alpha_0} \int_{\beta_0} &g_{do}^3 \epsilon(r)\, \delta(t-\tau) \\
    &\ln g_{sd} \mathcal{R}\left(E_o|\mu_e,\Gamma,g_{sd}E_{cut} \right) d\alpha_0 , \,d\beta_0
\end{split}
\end{equation}

\begin{equation}
\begin{split}
    w_2(E_o, t) = \int_{\alpha_0} \int_{\beta_0} &g_{do}^3 \epsilon(r)\, \delta(t-\tau)\\ &\frac{\partial\mathcal{R}\left(E_o|\mu_e,\Gamma,g_{sd}E_{cut}\right)}{\partial\Gamma} d\alpha_0 \,d\beta_0 .
\end{split}
\end{equation}
In the Fourier transform of these expressions the delta function $(\delta(t-\tau))$ becomes $e^{2\pi\tau\nu}$.

\section{Ionization calculation}
\label{app:ion_calc}
In order to estimate the upper limit of the ionisation parameter, we do not consider the time dependence of the direct emission and we use the definitions of Ingram et al (2019). Let us start from the definition of the ionisation parameter in the disc at radial coordinate $r$ \citep[see e.g. Eq.~10 in][]{Garcia2013}
\begin{equation}
    \xi(r) = 4\pi  \frac{F_{d,in} (r)}{n_e(r)},
\end{equation}
where 
\begin{equation}
    F_{d,in}(r) \equiv \frac{C}{2} \epsilon(r) \int_{13.6 {\rm eV}}^{13.6 {\rm keV}} E_d^{1-\Gamma} {\rm e}^{-E_d/(g_{sd}E_{\rm cut})} dE_d,
    \label{eq:app_flux_in_disc}
\end{equation}
is the incident flux from the source to the disc in the disc restframe. 
However, the model for the direct emission that we use to fit to the data is in the observer restframe
\begin{equation}
    F_{o}(E_o) = A\, l\, g^{\Gamma}_{so} E_o^{1-\Gamma} {\rm e}^{-E_o/(g_{so}E_{\rm cut})},
\end{equation}
where the normalization $A$ of \textsc{reltrans} model (the \textsc{xspec} model parameter `norm') absorbes the distance $(D)$ to the source and the normalisation of the direct emission with $A = C/(4 \pi D^2)$. Here, $E_{\rm cut}$ is defined in the source restframe. We can define $\mathcal{F}$ 
\begin{equation}
    \mathcal{F} \equiv A \int_{13.6 {\rm eV}}^{13.6 {\rm keV}} E_o^{1-\Gamma} {\rm e}^{-E_o/(g_{so}E_{\rm cut})} dE_o,
\end{equation}
because it comes directly from the model that we fit to the data. In order to express the ionisation parameter in terms of observed flux we can start from Eq.~\ref{eq:app_flux_in_disc} and apply coordinate transforms from $E_d$ to $E_s$ and then from $E_s$ to $E_o$ using the blueshift definitions $g_{so} = E_o/E_s$ and $g_{sd} = E_d/E_s$. 
We obtain 
\begin{equation}
    F_{d,in}(r) = 2 \pi \,D^2 \,\epsilon(r) \,g_{sd}^{2-\Gamma} \, g_{so}^{\Gamma - 2} \, \mathcal{F}.
\end{equation}
The emissivity as defined in Ingram et al (2019) is in units of $R_g^{-2}$. Keeping this definition, we can express the ionisation parameter in units of $[erg\, cm \, s^{-1}]$ as

\begin{equation}
     \xi(r)= \frac{8 \pi \,D_{\rm{cm}}^2 }{n_{e,\rm{cm}^{-3}}}\,\frac{\epsilon(r) \,g_{sd}^{2-\Gamma} \, g_{so}^{\Gamma - 2}}{R^2_{g,\rm{cm}}} \, \mathcal{F}_{\rm{erg}\, \rm{cm}^{-2}\, \rm{s}^{-1}}.
    \label{eq:app_xi}
\end{equation}
Finally, Eq~\ref{eq:app_xi} assumes isotropic radiation from the source. We should multiply by the boost parameter $(1/\mathcal{B})$. Eq~\ref{eq:ion_calc} in the text follows from subbing in sensible values for the parameters.

\section{MCMC analysis}
\label{app:MCMC}
In order to further probe the parameter space we run multiple MCMC simulations for Model~3 using the Goodman-Weare algorithm.
We run 4 chains using the \textsc{xspec} routine \textsc{chain}. The total length is $10^5$ for the first 3 and $2\times 10^5$ for the final chain. We use $200$, $400$, $800$ and $200$ walkers for the 4 chains.
All the walkers start from the best fit of Model~3 and we set the burn-in period to $4000$ steps in the first three chains and $2000$ steps for the last one. 
We test the length of the burn-in period with the Geweke convergence measure, finding this to be in the range $-0.2$ to $0.2$
for each parameter in each chain\footnote{The Geweke convergence measure compares two intervals of the chain, one shortly after the burn-in period and one towards the end of the chain. For each parameter, Geweke's statistic measures the difference between the mean parameter value in these two intervals, and therefore if convergence is achieved it will have an expectation value of zero and follow the standard normal distribution (\citealt{Geweke1992}).}, indicating that convergence has been achieved.

We jointly fit $21$ spectra (real and imaginary parts of the complex covariance for $10$ frequency ranges plus the time-average energy spectrum) so we have $51$ free parameters. The continuum phase ($\phi_A$ and $\phi_B$) and amplitude ($\alpha$ and $\gamma$) parameters are tied between real and imaginary parts for a given frequency, but all depend on frequency. They therefore contribute $40$ free parameters. The normalization of the time-average spectrum (which is effectively $\alpha(\nu=0)$), and the hydrogen column density of the absorption model contribute a further two free parameters. This leaves $9$ remaining parameters: $\Gamma$, $E_{\rm cut}$ and the 7 key parameters that we explore in detail: $M$, $1/\mathcal{B}$ ($\mathrm{Boost}$ in the model), $A_{\rm Fe}$, $\log\xi$, $r_{\rm in}$, Incl and $h$.

\begin{figure*}
	\includegraphics[width=\textwidth]{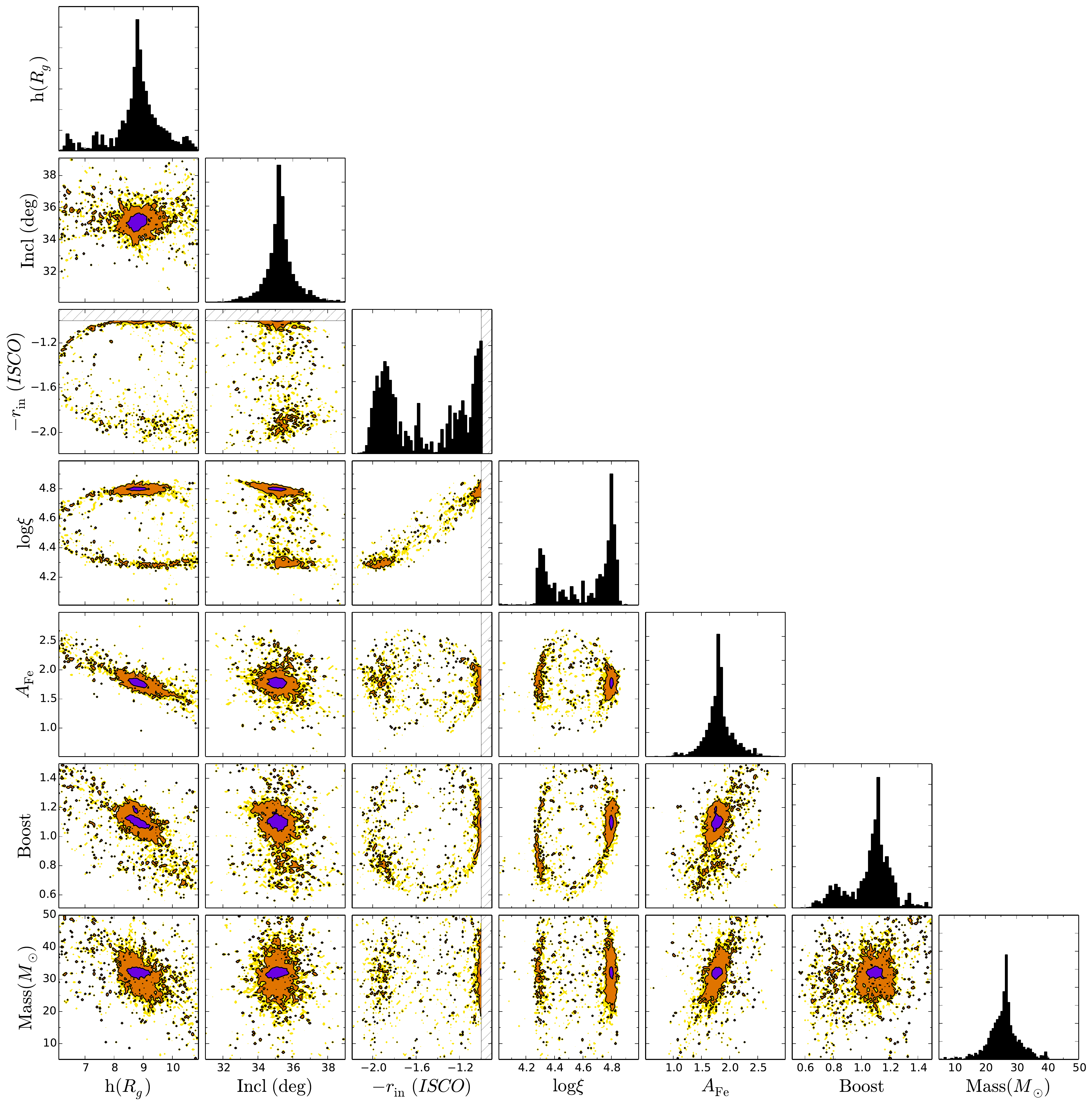}
    \caption{Output distributions from the MCMC simulation of Model~3. The purple, orange and yellow regions are $1\sigma$, $2\sigma $ and $3\sigma$ contours respectively. The inner radius is presented as negative following the \textsc{reltrans} convention and hatched regions indicate forbidden values. The y-axes for the histograms are in arbitrary units. }
    \label{fig:corner}
\end{figure*}

Fig.~\ref{fig:corner} is the integrated probability distribution exploring the correlations between these 7 key parameters using the MCMC. The inner radius is negative in order for it to be expressed in units of ISCOs in the \textsc{reltrans} model (following the \textsc{relxill} convention). The inner radius and peak ionization parameter appear to have bi-modal distributions, and anti-correlate with one another. The correlation between mass and peak ionization parameter revealed by the contour plot in Fig.~\ref{fig:2Dcontour} does not show up here, since the MCMC does not explore the high $\chi^2$ regions of parameter space as comprehensively as a brute-force grid search. For each parameter that we consider in the plot, the Rubin-Gelman convergence measure (\citealt{Gelman1992}) for the first three chains\footnote{The Rubin-Gelman convergence measure only compares chains of the same length} is $< 1.05$ (with values close to unity indicating convergence). We find that integrated probability distribution plots computed for each of the 4 individual chains all look very similar to Fig.~\ref{fig:corner}, indicating that the 4 chains sample similar distributions. In particular, this shows that the bi-modal behaviour of the $r_{\rm in}$ and $\log\xi$ parameters does not result from different chains returning different distributions.

\bsp	
\label{lastpage}
\end{document}